\documentclass[final,5p,authoryear,10pt]{elsarticle}

\usepackage{amssymb}
\usepackage{amsthm}
\usepackage{xfrac}
\usepackage{float}
\usepackage{hyperref}
\usepackage{fancyvrb}
\usepackage{color}
\usepackage{footnote} 
\usepackage[usenames,dvipsnames,svgnames,table]{xcolor}
\usepackage{subfig}

\newcommand\keff{k$_{\mathrm{eff}}$}

\newcommand\nubar[1]{$\overline{\nu}_\mathrm{#1}$}
\newcommand\ura{$^{235}$U}
\newcommand{\plut}{$^{239}$Pu}
\newcommand{\comment}[1]{}
\newcommand{\scale}[0]{SCALE~6.0}
\newcommand{\efb}{ENDF/B-VII.1}
\newcommand{\ef}{ENDF-6}
\newcommand{\exercise}{Exercise~I-1b}

\journal{Annals of Nuclear Energy}

\begin{document}

\begin{frontmatter}



\title{Comparison of nuclear data uncertainty propagation methodologies for PWR burn-up simulations}

\author[DIN]{C.J. Diez}
\ead{cj.diez@upm.es}

\author[AREVA1]{O. Buss}
\author[AREVA1]{A. Hoefer}
\author[AREVA2]{D. Porsch}
\author[DIN,DENIM]{O. Cabellos}

\address[DIN]{Universidad Polit\'ecnica de Madrid UPM \\
				Escuela T\'ecnica Superior de Ingenieros Industriales -- Dpto. de Ingenier\'ia Nuclear\\
        Jos\'e Guitierrez Abascal 2, 28006, Madrid, Spain\\}
\address[DENIM]{Universidad Polit\'ecnica de Madrid UPM \\
				Escuela T\'ecnica Superior de Ingenieros Industriales -- Instituto de Fusi\'on Nuclear\\
        Jos\'e Guitierrez Abascal 2, 28006, Madrid, Spain\\}

\address[AREVA1]{AREVA~GmbH -- Dept.~Radiology \& Criticality\\
    Kaiserleistrasse 29, 63067 Offenbach am Main, Germany\\}
\address[AREVA2]{AREVA~GmbH -- Dept.~Neutronics\\
        Paul-Gossen-Strasse 100, 91052 Erlangen, Germany\\}

\begin{abstract}
Several methodologies using different levels of approximations have been developed for propagating nuclear data uncertainties in nuclear burn-up simulations. Most methods fall into the two broad classes of Monte Carlo approaches, which are exact apart from statistical uncertainties but require additional computation time, and first order perturbation theory approaches, which are efficient for not too large numbers of considered response functions but only applicable for sufficiently small nuclear data uncertainties. Some methods neglect isotopic composition uncertainties induced by the depletion steps of the simulations, others neglect neutron flux uncertainties, and the accuracy of a given approximation is often very hard to quantify. In order to get a better sense of the impact of different approximations, this work aims to compare results obtained based on different approximate methodologies with an exact method, namely the NUDUNA Monte Carlo based approach developed by AREVA GmbH. In addition, the impact 
of different covariance data is studied by comparing two of the 
presently most complete nuclear data covariance libraries (\efb{} and \scale{}), which reveals a high dependency of the uncertainty estimates on the 
source of covariance data. 
The burn-up benchmark \exercise{} proposed by the OECD expert group 
``Benchmarks for Uncertainty Analysis in Modeling (UAM) for the Design, Operation and Safety Analysis of LWRs''
is studied as an example application. The burn-up simulations are performed with the \scale{} tool suite.
\end{abstract}


\end{frontmatter}


\newpage

\section{Introduction}
\label{sec:intro}
Nuclear transport and depletion codes are the work\-horses in the field of nuclear safety,
in reactor core design, and for radiation shielding analyzes. Such codes describe the propagation of sub-atomic particles, e.g.~neutrons, in matter and simulate the induced change in the material composition, e.g., due to neutron capture, fission, and radioactive decay processes. 

A vast variety of experimental and model input data is required in order to parametrize the interaction of sub-atomic particles with the surrounding matter  and the induced change in the material composition. Such nuclear data are usually provided in a standardized format, like \ef{}~\citep{endf6_format}, by nuclear data evaluation groups who release them in the form of nuclear data libraries, such as \efb{}~\citep{endfb71}, {JEFF-3.2~\citep{jeff32}} or JENDL-4~\citep{jendl4}. These nuclear data are then processed, filtered, compressed, and reformatted in order to serve as input for specific nuclear transport and depletion codes such as SCALE~\citep{scale60}, ACAB \citep{acab_manual}, and FISPACT-II~\citep{fispact2}.

The nuclear input data have limited accuracy which arises both from limited measurement precision and modeling uncertainties, e.g., in regions where insufficient experimental data are available. Hence, it is the objective of the nuclear data community to provide uncertainty estimates that truly reflect {the} degree of confidence in the nuclear data, and the last years have witnessed great progress in this field.

Besides the fundamental need of guaranteeing conservatism of safety assessments, there is another major motivation for studying the impact of nuclear data uncertainties. That is replacing very penalizing assumptions in safety assessments with accurate uncertainty estimates, which will result in an improved economic competitiveness. 

Over the last decades, several methodologies have been implemented in code packages to study the impact of nuclear data uncertainties on nuclear transport and depletion problems, such as TSUNAMI, TSURFER, and SAMPLER developed by ORNL~\citep{scale60}, XSUSA by GRS~\citep{xsusa}, RIB by CEA~\citep{rib_tool}, TMC by NRG~\citep{tmc1}, FISPACT-II by CCFE~\citep{fispact2}, NUSS by PSI~\citep{PSI,PSI_nuss} developed for CASMO-5M~\citep{CASMO5} and MCNPX~\citep{mcnpx}, the Kiwi package by LLNL~\citep{kiwi_tool}, the so-called Hybrid Method~\citep{hybrid_method} based on ACAB~\citep{acab_manual}, promoted by Universidad \linebreak Polit\'ecnica de Madrid, and the NUDUNA package developed by AREVA GmbH~\citep{nuduna}. The methods can be divided into first order perturbation theory based methods (TSUNAMI, TSURFER, RIB, FISPACT-II) and Monte Carlo sampling methods (SAMPLER, XSUSA,\linebreak TMC, Kiwi, Hybrid Method, NUSS, NUDUNA). 

A first order perturbation theory method typically requires little extra computation time compared to an 
analysis without uncertainty quantification. E.g., an adjoint-based sensitivity analysis involving the application of the so-called sandwich rule~\citep{cacuci} requires only as many importance calculations as there are response functions under consideration. Hence, the adjoint-based approach is efficient for a sufficiently small number of response functions. 
Monte Carlo methods, on the contrary, are not limited to small nuclear data uncertainties because they do not involve any series expansions and, consequently, completely take into account nonlinear effects. Furthermore, they are easy to implement without any major code modification. However, it can be seen as a disadvantage that they usually require elevated CPU time compared to first order approximation approaches, although there are application cases where the increase in CPU time can be rendered negligible \citep{fastGRS2012,fastTMC2013}.
Methods also differ by their nuclear data uncertainty input. E.g., the TMC method is directly based on microscopic measurements, while up to now all other methods are based on data from nuclear data evaluations. 

\begin{figure*}[tb]
  \centering
  \resizebox{1.4\columnwidth}{!}{\rotatebox{0}{\includegraphics[]{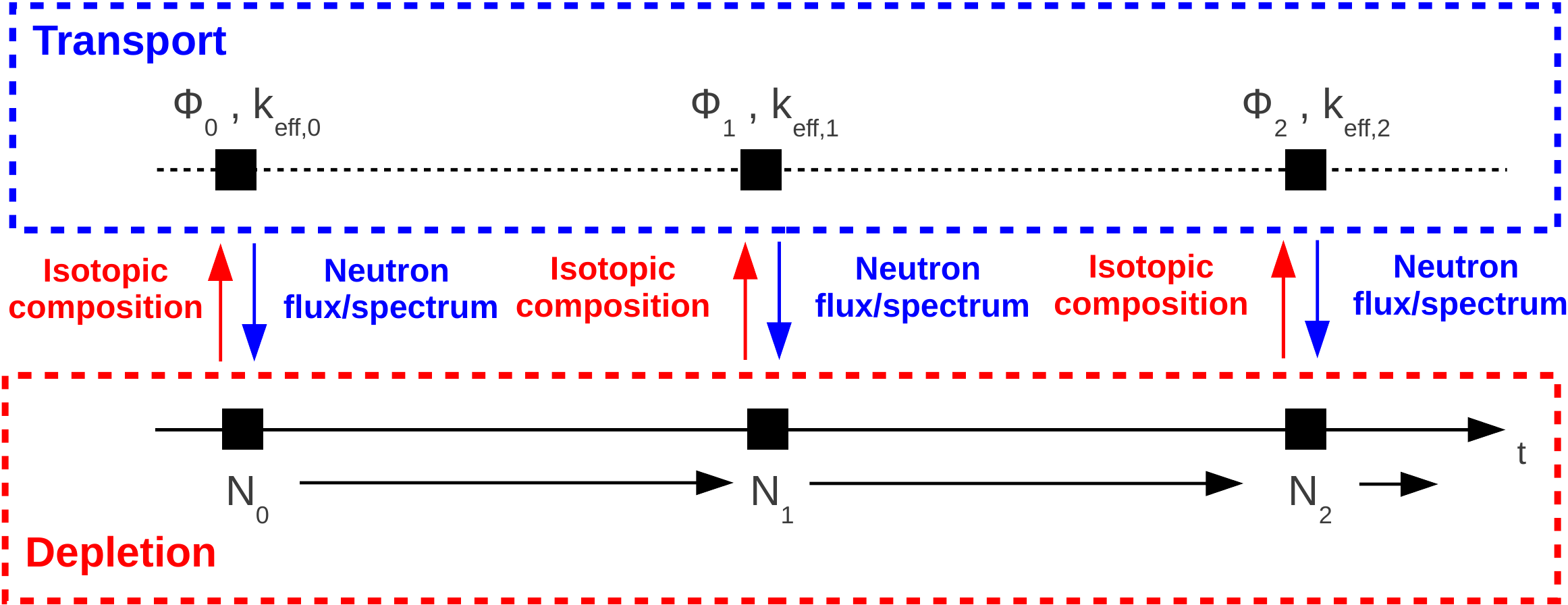}}}
  \caption{Calculation scheme for burn-up simulation: coupling of transport and depletion codes.}
  \label{fig:burn-up-scheme}
\end{figure*}

When simulating the burn-up of nuclear fuel in a reactor, the problem is usually divided into evaluation of the neutron flux and depletion of the fuel. Fig.~\ref{fig:burn-up-scheme} shows a typical scheme for such a simulation, which includes the coupling between a transport code, which estimates the energy-dependent neutron flux $\Phi$, and a depletion code, which describes the change of the material compositions. Typically, depletion codes such as MONTEBURNS \citep{monteburns}, TRITON \citep{scale60}, and SERPENT \citep{serpent} solve the underlying differential equations with the help of so-called predictor-corrector algorithms.

Nuclear data uncertainties affect 
both transport and depletion codes. {Uncertainties in cross sections, angular distributions, neutron multiplicities and fission spectra imply uncertainties in the neutron flux and the neutron energy spectrum.} This flux and spectrum uncertainty has to be propagated during the depletion analysis which is, additionally, explicitly affected by cross section uncertainties and by decay data and fission yield data uncertainties. After a depletion step, the updated material composition carries an uncertainty which then has to be propagated to the following transport step.

In a full Monte Carlo approach, the nuclear data are sampled at the beginning of the burn-up simulation, then the cycle of transport and depletion is executed for one Monte Carlo sample, then the next sample is generated and used for a next simulation. The statistics of all simulations finally yields the desired uncertainty statements. An approximate way to evaluate the burn-up uncertainty is given by the Hybrid Method~\citep{hybrid_method}. With this method, flux and spectrum uncertainties are neglected and only the uncertainties in the depletion step are propagated. This has the advantage that it requires performing the transport calculation only once for nominal nuclear data and, consequently, avoids the costly repetitions of the transport code updates.

This work focuses on the impact of different approximations to the propagation of nuclear data uncertainties. As outlined above, the first task will be
to check the performance of the first order approximation. The next question that has to be addressed is what is the loss in accuracy related to neglecting either the uncertainties in the transport step or the depletion step. Additionally, the impact of different nuclear data covariance library input is studied by performing the same Uncertainty Quantification (UQ) analysis with two different inputs. For this we have chosen two of the presently most complete nuclear data covariance libraries: \efb{} and \scale{}. The reference to the approximate methods will be the NUDUNA method which is a Monte Carlo based approach that propagates the uncertainties in a complete fashion. As an example application, we study the burn-up benchmark \exercise{} proposed by the OECD expert group  ``Benchmarks for Uncertainty Analysis in Modeling (UAM) for the Design, Operation and Safety Analysis of LWRs''. The 
burn-up simulations are performed with the \scale{} tool suite.

{Note that uncertainties induced by so-called technological parameters, e.g.~system dimensions and initial material compositions, and the operating history of a given fuel element, could lead to uncertainty estimations of the same magnitude as the estimations from nuclear data uncertainties~\citep{nrg_tmc_pwr_burnup_pincell_report}, but their discussion is beyond the scope of this paper.}

This paper is structured as follows. First, the UAM benchmark and the nuclear covariance inputs are discussed. Next, the NUDUNA method is introduced. The main part addresses the impact of different approximations and of different state-of-the-art covariance inputs. Finally, a full UQ analysis for the considered UAM benchmark is presented.

\section{UAM benchmark exercise on nuclear data uncertainties in a pin-cell burn-up calculation}
\label{sec:def_problem}
\begin{figure*}[tb]
  \centering
  \resizebox{1.6\columnwidth}{!}{\rotatebox{0}{\includegraphics[]{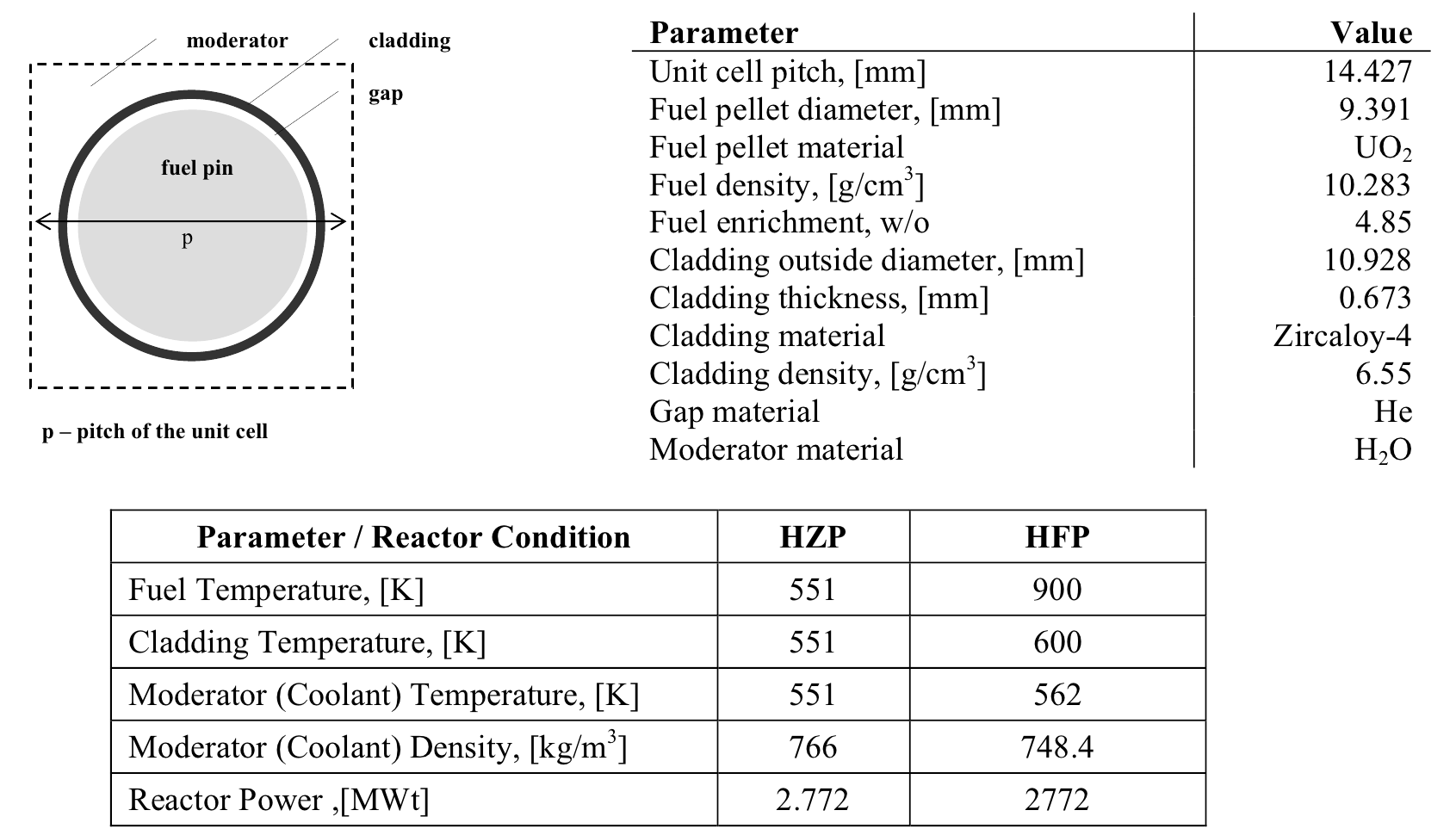}}}
  \caption{Specifications of the UAM \exercise{} modeling a PWR pin-cell (provided by \cite{uam_specification}).}
  \label{pin-cell-spc}
\end{figure*}

The UAM Benchmark \exercise{} is described in ~\cite[Appendix VIII]{uam_specification}. It consists in performing a UQ study on a typical pressurized water reactor~(PWR) pin-cell burn-up calculation. The main specifications are summarized in Fig.~\ref{pin-cell-spc}. Here we only address  Hot Full Power (HFP) conditions, where the average power density is $33.58\,\mathrm{W/gU}$, and is kept constant during the whole burn-up period of $1825$ days. Thus a final burn-up of 61.28~GWd/MTU is achieved.

{The burn-up calculations are performed with the TRITON sequence from \scale{}, which uses ORIGEN as the depletion code.}
Throughout the burn-up process, the neutron multiplication factor \keff{} and the concentrations of the 
isotopes given in Table~\ref{tab:list_iso_analysed} are followed and analyzed.

\begin{table}[tb]
  \centering
  \caption{List of isotopes whose concentrations are followed throughout the burn-up process.}
  \label{tab:list_iso_analysed}
  \footnotesize
  \begin{tabular}{cc}
    \hline
    \hline
Light isotopes &  $^{16}$O, $^{90}$Sr, $^{95}$Mo, $^{99}$Tc, $^{109}$Ag, $^{103}$Rh \\
	{and}    &  $^{106}$Ru, $^{133}$Cs, $^{134}$Cs, $^{137}$Cs, $^{135}$I, $^{135}$Xe \\
{fission products} &  $^{139}$La, $^{151}$Eu, $^{153}$Eu, $^{154}$Eu, $^{155}$Eu, $^{155}$Gd \\
	       &  $^{156}$Gd, $^{143}$Nd, $^{144}$Nd, $^{145}$Nd, $^{146}$Nd, $^{148}$Nd \\
	       &  $^{147}$Sm, $^{149}$Sm, $^{151}$Sm, $^{147}$Pm \\
\hline
Heavy isotopes &  $^{234}$U, $^{235}$U, $^{236}$U, $^{238}$U, $^{237}$Np, $^{238}$Pu \\
	       &  $^{239}$Pu, $^{240}$Pu, $^{241}$Pu, $^{242}$Pu, $^{243}$Am, $^{244}$Cm \\
    \hline
    \hline
  \end{tabular}
\end{table}


\section{Nuclear data covariance input}
\label{sec:why_scale60_endfb71}
Nuclear data uncertainty quantification is still a developing field, which is demonstrated by the fact that covariance data sometimes significantly change even from one release of a nuclear data library to the other, e.g.~when switching from ENDF/B-VII.0~\citep{endfb70} to \efb{}{, as presented by~\cite{endfb70_enfb71_cov_changes}}. Since uncertainty estimations from UQ studies reflect the uncertainty data in nuclear data libraries, new developments in nuclear data uncertainty quantification will inevitably lead to different UQ results.

{In order to study the impact of different covariance data input, we decided to compare here two of the currently most complete covariance libraries with regard to the data which are important for criticality and burn-up calculations~\citep{nrg_tmc_pwr_burnup_pincell,oscar_stni_2013}: neutron cross section (including resonant and non-resonant region), neutron multiplicity and neutron emission spectrum uncertainties.
Angular distribution uncertainties have not been identified as relevant for this tpye of analysis~\citep{nrg_tmc_pwr_burnup_pincell_report}, and are not separately investigated in this work.
The selected covariance libraries are the \scale{} covariance library~\citep{scale60} and the covariances provided in \efb{}~\citep{endfb71}. The \scale{} covariances have also been selected by the UAM expert group as a reference library for the benchmark exercise after reviewing the state-of-the-art of cross section covariance data~\citep[Sec.2]{uam_specification}. At the time of that review, the \scale{} covariance library was the most complete and up-to-date compilation, and only thereafter a considerable amount of additional covariance information was released as part of the \efb{}, JENDL-4.0, and JEFF 3.2 evaluations.}

The SCALE 6.0 covariance library consists of two parts:
\begin{itemize}
  \item ``High fidelity'' uncertainty data taken 
from ENDF/B-VI.8~\citep{endfb68}, ENDF/B-VII.0, a pre-release of \efb{} and JENDL-3.3~\citep{jendl33}
for more than 50 nuclides, including the most important ones for LWR applications.
  \item ``Low fidelity'' uncertainty data, estimated independently of a specific data evaluation, retrieved from
a collaborative project aimed to provide covariances over the energy range from 10$^{-5}$ eV to 20 MeV for
materials without covariances in ENDF/B-VII.0.\linebreak These data were called the ``BLO'' [BNL-LANL-ORNL] uncertainty data~\citep{blo_unc_lib}.
\end{itemize}
It provides uncertainties for a total of 401 nuclides in the form of covariance matrices in 44 energy groups
for cross sections, fission neutron multiplicities \nubar{} and fission neutron energy spectra $\chi$. It also includes covariance data between reactions of different isotopes. Various UQ methodologies use the \scale{} uncertainty data as input, such as TSUNAMI, XSUSA, and PSI-NUSS.

\efb{} now also includes comprehensive covariance data and collects efforts of several projects performed between 2006-2011 on nuclear data uncertainties:
\begin{itemize}
  \item BOLNA. A covariance library created by five laboratories (BOLNA = Brookhaven-Oak Ridge-Los
	Alamos-NRG Petten-Argonne) for the purposes of the international project WPEC Subgroup 26~\citep{wpec_sg_26}.
	The library represents an ad hoc collection of covariances not tied to any specific evaluated nuclear data library.
  \item ``Low fidelity'' uncertainty data, provided within BLO.
  \item COMMARA-2.0~\citep{commara20}. A library produced by a BNL-LANL collaboration during 2008-2011.
	It constitutes the backbone of \efb{} covariances, once adapted to the central values proposed for \efb{}.
\end{itemize}
Apart from that, re-evaluations were also made for \linebreak \efb{}, but only for a few nuclides. \efb{} covariance data have been supplemented by covariance evaluations from JENDL-4.0, providing complete covariance data for a total of 190 nuclides. {So, it provides uncertainty data for neutron multiplicities, fission neutron spectra, cross sections and angular distributions.
Uncertainties for the unresolved resonance region (URR) are typically included in the cross section uncertainties (i.e. in file~33 of an ENDF-6 tape), and only for very few isotopes there are detailed URR parameter uncertainties given (e.g.~$^{232}$Th).}
JEFF-3.2 was released early in 2014 and includes covariance data for more isotopes than \efb{}, but lacks data for important fuel isotopes. E.g., the \plut{} cross section file does not provide any non-resonance cross section uncertainties. Thus we have decided to perform our study based on \efb{}, which represents for our field of application the most up-to-date covariance input provided by a nuclear data evaluation group.

\section{NUDUNA program package for nuclear data uncertainty analysis}
\label{sec:NudunaTool}
The NUDUNA (NUclear Data UNcertainty Analysis) program package has been developed within the last four years by AREVA GmbH. It will be applied as reference method when assessing the impact of different approximations.

\begin{figure*}[tb]
  \centering
  \resizebox{1.4\columnwidth}{!}{\rotatebox{0}{\includegraphics[]{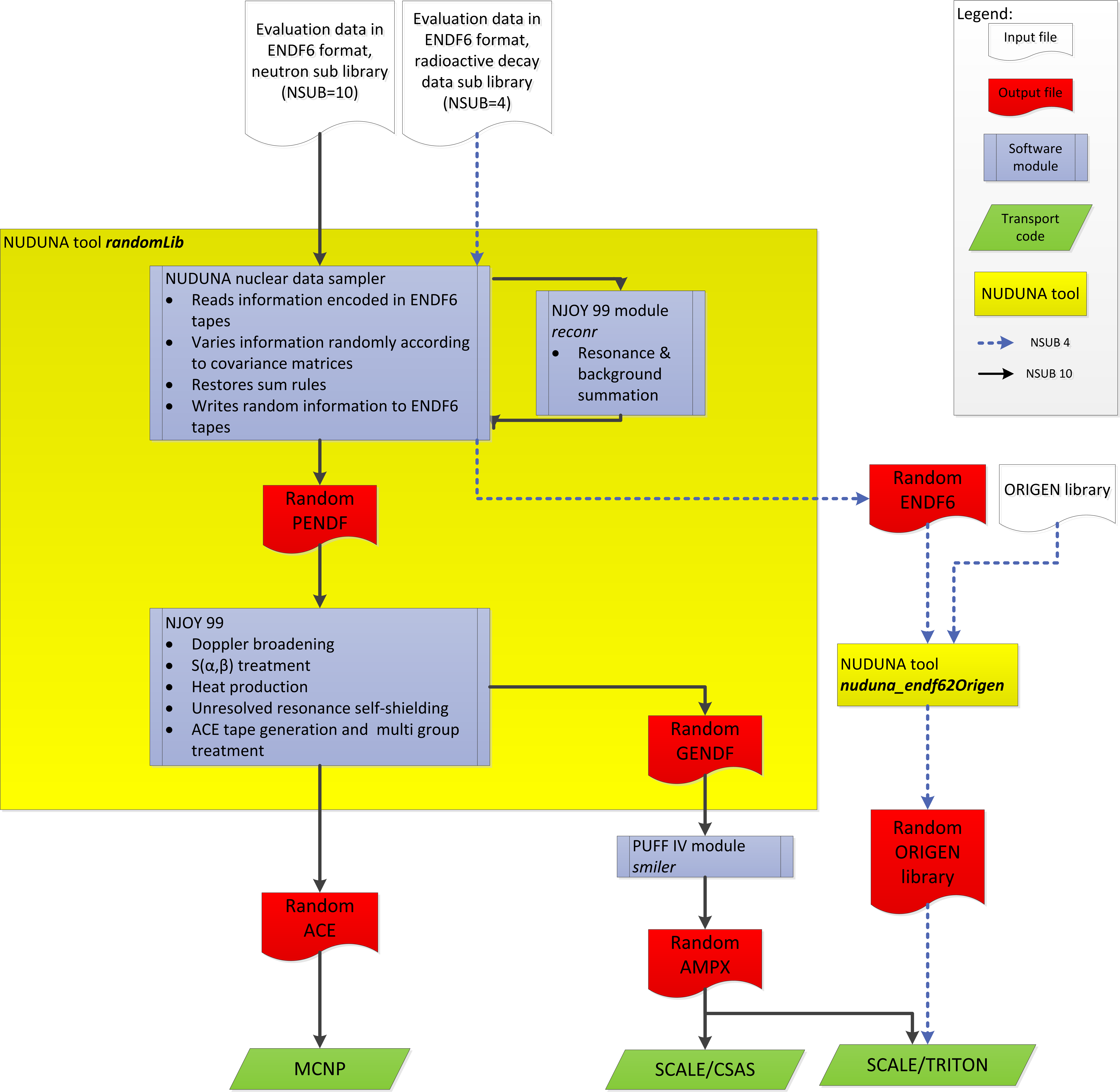}}}
  \caption{Sampling of nuclear data input libraries with NUDUNA.}
  \label{fig:nuduna_mc_scheme}
\end{figure*}

NUDUNA provides full Monte Carlo sampling of the nuclear data inputs for transport and depletion calculations. Given such a tool, 
one can draw random samples of nuclear data and perform a separate transport and/or depletion calculation for each random sample.
Then, the distribution of the individual results corresponding to the different random inputs can be analyzed, and uncertainty estimates can be deduced. The flowchart of the NUDUNA random sampling procedure is depicted in Fig.~\ref{fig:nuduna_mc_scheme}.

NUDUNA takes as input the information provided by nuclear data evaluations in the standardized \ef{}~format~\citep{endf6_format}.
Such \ef{} formatted files do not only supply best estimate values of cross sections, fission yields, decay data etc., but also their covariances.
NUDUNA is capable of drawing random samples of nuclear data based on this covariance information.

At present, NUDUNA {is validated} for automatic compilation of random input files for the MCNP5 code \citep{mcnp5} and the \scale{} tool suite, i.e.~it can be applied, amongst others, to perform random simulations with the TRITON depletion sequence from \scale. {However, the NUDUNA output files could also be used as input for other codes, e.g. Serpent or MCNP6.}

\subsection{Nuclear data stored in \ef{} format}
\begin{figure}[tb]
  \centering
  \resizebox{1.0\columnwidth}{!}{\rotatebox{0}{\includegraphics[]{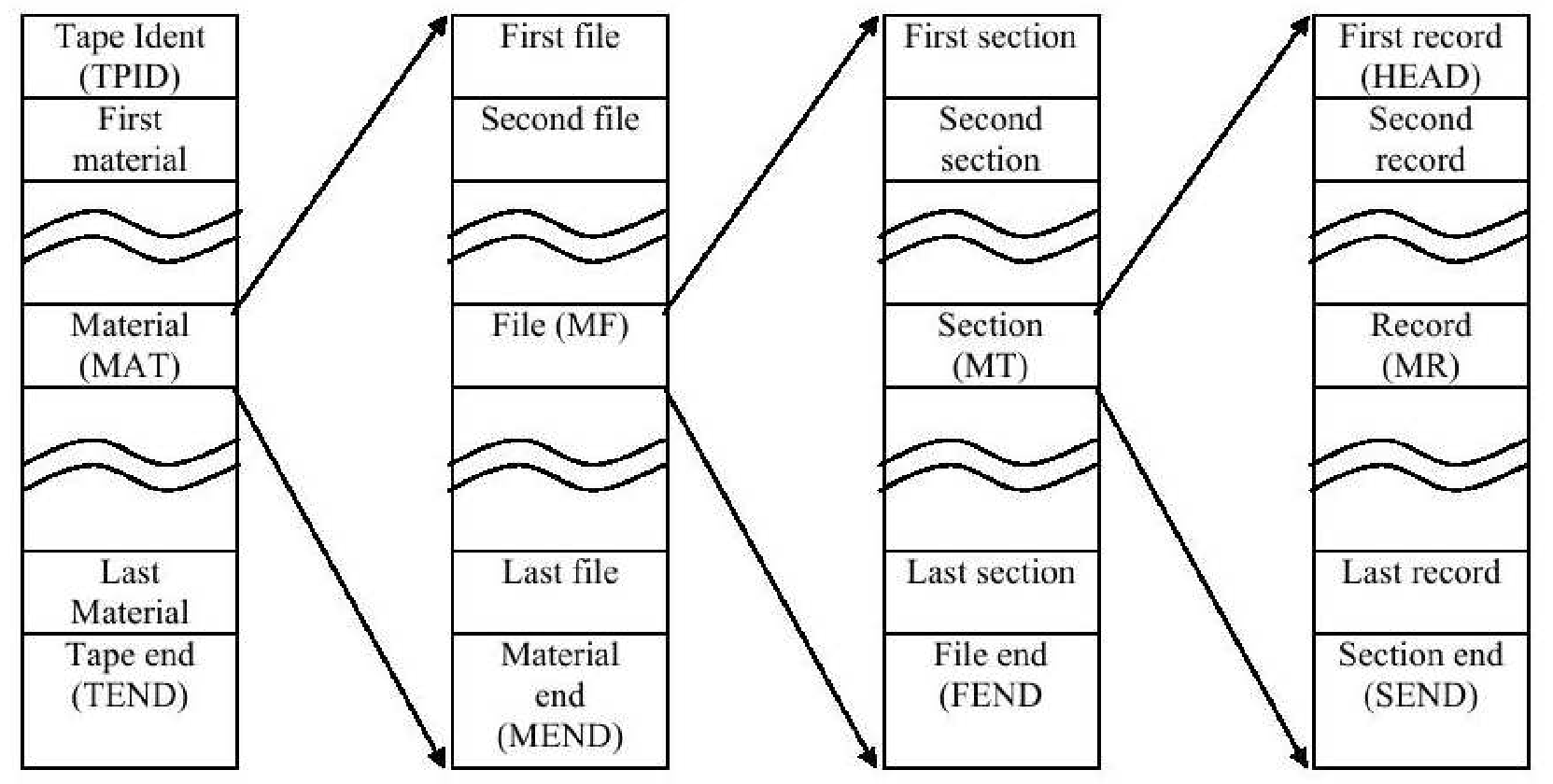}}}
  \caption{Structure of an \ef{} type tape (taken from ~\citep{endf6_format}).}
  \label{fig:ENDF6_structure}
\end{figure}

The structure of an \ef{} tape is hierarchical and sketched in Fig.~\ref{fig:ENDF6_structure}.
Each tape contains a library which may have several sections representing different materials (MAT).
The library type defines the incoming projectile:
there are libraries for photon, neutron, and proton induced events and for decay data.
Each material section is structured into several so-called \textit{Files}.
The most relevant files for NUDUNA are Files~1-8 and 31-35.
File~1  contains general information and the multiplicities of neutrons for prompt and delayed fission reactions,
File~2  contains resonance parameters,
File~3  contains non-resonance cross sections,
Files~4-6  are used to store energy and angular distributions of final state particles,
File~7 contains thermal scattering data - S($\alpha$,$\beta$),
File~8  provides radioactive decay data and fission product yields.
Files~31-35  store the covariance information for Files 1-5.
Each file contains sections providing information on a specific reaction type,
and the sections themselves are structured in several records.

\subsection{Random sampling of nuclear data}
\label{sec:gen_random_ENDF6}
NUDUNA is capable of reading nuclear data encoded in \ef{}~format and of random sampling the following data according to their covariance information:
\begin{itemize}
  \item average fission neutron multiplicities \nubar{} (File~1),
  \item resonance parameters (File~2),
  \item cross sections (File~3),
  \item angular distributions (File~4),
  \item decay data (File~8, Section~457).
\end{itemize}
The \ef{}~format provides only nominal values and covariances, but no further information on the distribution of the data. NUDUNA users can choose between normal and log-normal probability distributions. In this paper, we consider only sampling based on normal distributions. The following paragraph details the sampling procedure.  At present, NUDUNA does not consider the covariance information between different nuclides, which usually leads to negligible uncertainty contributions if the thermal energy region of the neutron spectrum is of importance, e.g.~for PWR reactors~\citep{oscar_stni_2013}.

{NUDUNA provides random PENDF (point-wise ENDF format) files which are then processed with NJOY and PUFF in order to generate random GENDF, ACE, and AMPX files (cf. Figure~\ref{fig:nuduna_mc_scheme}), which serve as inputs for transport codes.}

\subsubsection{Fission neutron multiplicity \nubar{}}
\label{sec:mf1}
The fission neutron multiplicity \nubar{}, i.e.~the average number of neutrons per fission, is stored in File~1, which includes information for the total ($\overline{\nu}_\text{t}$), prompt ($\overline{\nu}_\text{p}$) and delayed ($\overline{\nu}_\text{d}$) fission neutron multiplicity. They are connected via
\begin{equation}
\label{eq:multiplicities}
 \overline{\nu}_\text{t}(E) = \overline{\nu}_\text{d}(E) + \overline{\nu}_\text{p}(E).
\end{equation}

The multiplicity data are sampled according to their covariances provided in File~31. After the sampling, the sum rule given in Eq.~\ref{eq:multiplicities} is checked and, if necessary, restored. If the \ef{} file provides no information on how this rule should be restored and if there is uncertainty information given for both prompt and total multiplicities, the total multiplicity is re-calculated as the sum of prompt and delayed multiplicities.

\subsubsection{Resonance parameters}
The resonance region is divided into the so-called Resolved Resonance Region (RRR)
and the Unresolved Resonance Region (URR). The parameters that describe the resonances
in both sub-regions and their uncertainties are given in File~2 and 32, respectively.

In the RRR regime, NUDUNA supports the most important formalisms (Reich-Moore, Single-level Breit-\linebreak Wigner and Multilevel Breit-Wigner). It randomizes all parameters according to their covariances and enforces the positivity bounds for the widths and energy parameters. {The uncertainties in the URR regime are predominantly treated as non-resonant cross section uncertainties within \efb{}, and the SCALE covariance library does not include explicit URR parameter uncertainties. So explicit URR parameter uncertainties are not considered in this work.}

\subsubsection{Non-resonance cross sections}
Non-resonance cross section data and their uncertainties are stored in Files 3 and 33, respectively. These cross sections have to be added to the resonance cross sections defined by the RRR and URR parameters. As stated in the \ef{}~format manual, the covariance information
provided in the cross section covariance File~33 applies to
the sum of non-resonance and resonance contributions. Thus, the resonance cross sections have to be reconstructed and added to the File~3 cross section before the random sampling can be performed.

The covariance data are given for energy ranges, and it is assumed that all points lying in the same energy range are completely correlated. Finally, sum rules have to be fulfilled. In principle, the \ef{} format provides with its LTY=0 sections a mechanism for defining how sum rules shall be restored. In case that this information is missing, the following procedure is applied:
\begin{itemize}
  \item If a cross section is given by the sum of others, e.g.~the total cross section,  and has no uncertainty information,
	this cross section is calculated using its sum rule.
  \item If there is covariance information for the sum and at least for one of the addends,
	the sum is evaluated as sum of the random draws of all addends.
  \item If there is uncertainty information for the sum but not for any of the addends,
	the addends are re-scaled in order to fulfill the sum rule.
\end{itemize}

\subsubsection{Angular distributions}
Angular distributions of final state particles and their uncertainties are stored in Files 4 and 34, respectively.
Usually, they are expressed as normalized probability distributions
given in Legendre representation. The Legendre coefficients are randomized, and the positivity of the distribution is enforced by rejecting samples that lead to negative distributions.

\subsubsection{Decay data}
Decay data are stored in File~8, Section 457 of an \ef{} radioactive decay data sub-library 4. The current \ef{}~format supports neither covariance matrices for branching ratios nor correlations between data of different isotopes, i.e. the numerical treatment is simple. NUDUNA samples both half-life values and branching ratios. For the branching ratios $\beta_i$ one has to enforce the constraint
	  \begin{equation}
	  \label{eq:branchingSum}
		\sum_i{\beta_i}=1\; .
		\end{equation}
This constraint implies correlations of the different branching ratios. However, \ef{} only includes their standard deviations but not their correlation matrix. For cases where there are two branching ratios, the standard deviations for the two channels have to be identical if the constraint of Eq.~\ref{eq:branchingSum} is taken into account in the uncertainty assessment. If the \ef{} file includes a ``0'' entry for one of two branching ratios, we replace it by the non-zero value of the other branching ratio. If two non-identical standard deviations are provided or if there are more than three decay channels, then the constraint of Eq.~\ref{eq:branchingSum} is imposed by means of Bayesian updating of the input uncertainties. 

The sampled decay data are checked for validity: branching ratios and half-lives have to be positive,
and the former cannot get larger than $1$. By default, NUDUNA assigns a 100\% uncertainty to every decay data entry for which no uncertainty information is provided.

\subsection{Converting \ef{} files into code-dependent format}
NUDUNA currently provides the capability of generating input ACE libraries for MCNP as well as AMPX and ORIGEN libraries for \scale{}. In this paper, we restrict ourselves to SCALE applications.

The AMPX format is a multi-group format, which is compiled by NUDUNA with the help of the NJOY~\citep{njoy} and PUFF~\citep{puff4} codes, as illustrated in Fig.~\ref{fig:nuduna_mc_scheme}. NJOY is applied to convert \ef{} files to group-wise \ef{} formatted tapes based on the 238-groups structure and collapsing spectra of SCALE~\citep{scale60} which is suitable for typical PWR reactor applications. Then PUFF is run to convert these files into AMPX files. As stated in Sec.~\ref{sec:gen_random_ENDF6}, the resonance and non-resonance cross sections have to be summed before the sampling of the cross sections. For this the NJOY module \textit{reconr} is applied.

The compilation of ORIGEN decay data libraries for \scale{} is also depicted in Fig.~\ref{fig:nuduna_mc_scheme}. Several limitations of the ORIGEN format have to be taken into account:
\begin{itemize}
  \item \ef{} format can handle multiple particle emission decay modes,
	while ORIGEN format can store only $\beta^-$+n.
	Thus, branching ratios of any multiple particle emission decay mode
	that involves at least a $\beta^-$+n are added to the $\beta^-$+n decay mode.
  \item \ef{} format can provide branching ratios to daughters in excited levels higher than the first,
	while ORIGEN can only handle decay modes to the first excited (metastable) state.
	Therefore, branching ratios to higher than the first excited (metastable) states of the daughter isotope
	are added to the branching ratio to the first excited (metastable) state.
  \item \ef{} format provides neutron emission decay modes (not related to $\beta^-$+n),
        while ORIGEN does not. Thus this decay mode is omitted in the conversion from \ef{} to ORIGEN format.
\end{itemize}


\section{Impact of approximations}
In the following, we are going to study the impact of {different} approximations to nuclear data uncertainty propagation:
\begin{itemize}
  \item first order approximation,
  \item neglecting number density uncertainties, and
	\item neglecting neutron flux and spectrum uncertainties.
\end{itemize}
Results obtained with approximate methods are compared to results obtained with NUDUNA which solves the problem in a complete manner based on Monte Carlo sampling and is free of the above approximations. These studies are based on \scale{} covariances, and, for the sake of simplicity, only \ura{} and \plut{} uncertainties are propagated, and covariances for total, elastic, (n,$\gamma$),
(n,2n), fission and inelastic cross section, plus correlations between elastic-fission, elastic-(n,$\gamma$) and (n,$\gamma$)-fission are included. Correlations between cross section reactions of different isotopes are not included,
as they have been shown to be irrelevant~\citep{oscar_stni_2013}.

\subsection{First order approximation}
\label{sec:effect_only_tran}
In first order approximation, the uncertainty of a desired response function is obtained by applying the so-called sandwich formula \citep[Sec.III.F]{cacuci}, which involves sensitivity coefficients and covariances. Current tools that apply first order approximation can tackle transport and depletion problems - however, a complete implementation that couples the two problems, as needed for the simulation of burn-up, is not yet available. \cite{oscar_stni_2013} has attempted to solved this issue, but the method still misses some effects. 

\begin{figure}[tb]
  \centering
  \subfloat[Due to $^{235}$U uncertainties]{\includegraphics[]{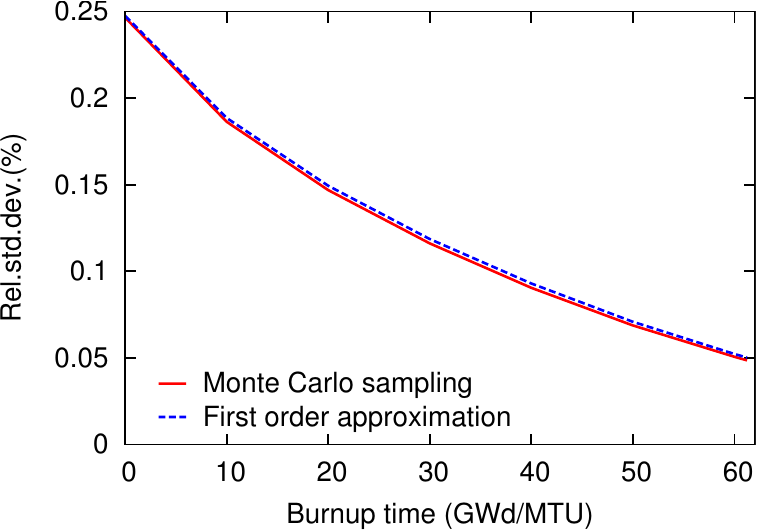}}\\
  \subfloat[Due to $^{239}$Pu uncertainties]{\includegraphics[]{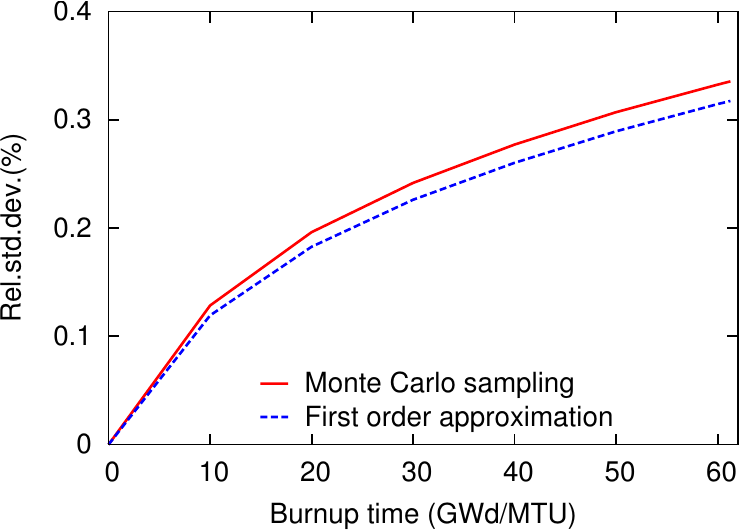}}
  \caption{\keff{} uncertainty due to transport calculation uncertainties at selected burn-up points based on \scale{} cross section uncertainties. The panels show results obtained with first order approximation (TSUNAMI) and Monte Carlo sampling (NUDUNA).}
  \label{fig:newt_tsunami_nuduna}
\end{figure}

One of the best known tools for uncertainty propagation in transport calculations is the TSUNAMI tool which is part of the SCALE package. It is based on first order approximation and can only estimate uncertainties of transport problems. Nevertheless, we can use it to study the performance of this approximation. For this we neglect the isotopic number density uncertainties, and perform at each burn-up step a TSUNAMI and a NUDUNA analysis for given inventory using identical \scale{} covariance input. Thus only the uncertainty due to flux and spectrum uncertainty will be obtained. Fig.~\ref{fig:newt_tsunami_nuduna} shows the predicted \keff{} uncertainty with TSUNAMI (data from \citep{oscar_stni_2013}) and NUDUNA, where the upper panel gives the uncertainty induced by $^{235}$U and the lower panel the one induced by $^{239}$Pu input data uncertainties. Good agreement between the TSUNAMI and NUDUNA results is found {for \keff{}}, which might suggest that a first order treatment is generally appropriate, at 
least for thermal light water systems. However, one has to keep in mind that nonlinear effects become more important for larger nuclear data uncertainties. Hence, first order results are expected to be less accurate for application cases sensitive to isotopes with large cross section uncertainties, such as the Gadolinium isotopes $^{155}$Gd and $^{157}$Gd.

\subsection{Neglecting the isotopic concentration uncertainties}
\label{sec:neglecting_n_unc}
NUDUNA is capable of propagating uncertainties \linebreak through the complete burn-up process and thus considers both isotopic and flux uncertainties. However, one can also perform a limited analysis where the isotopic concentration uncertainties are neglected (cf.~previous section). In the following, we are going to study the impact of such an approximation with special focus on the uncertainty contributions coming from cross section and neutron multiplicity data.

\subsubsection{Impact of cross section uncertainties}
\label{sec:neglecting_n_unc_xs}
\begin{figure}[tb]
  \centering
  \subfloat[\keff{} uncertainty]{\includegraphics[]{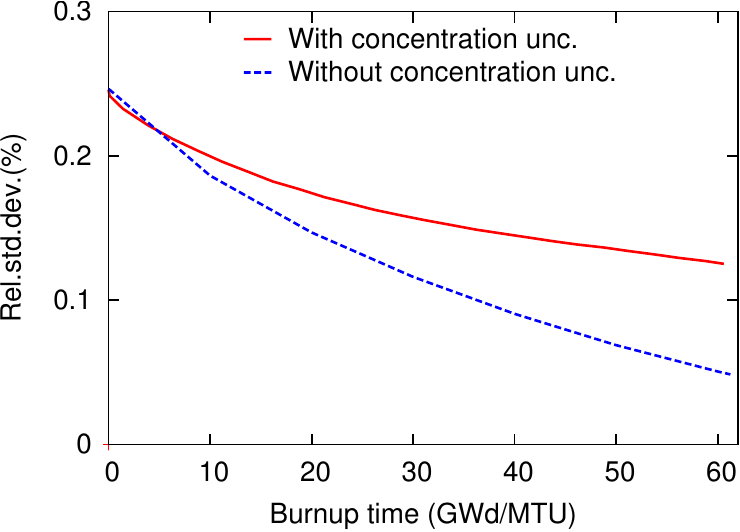}}\\
  \subfloat[Concentrations and their uncertainties]{\includegraphics[]{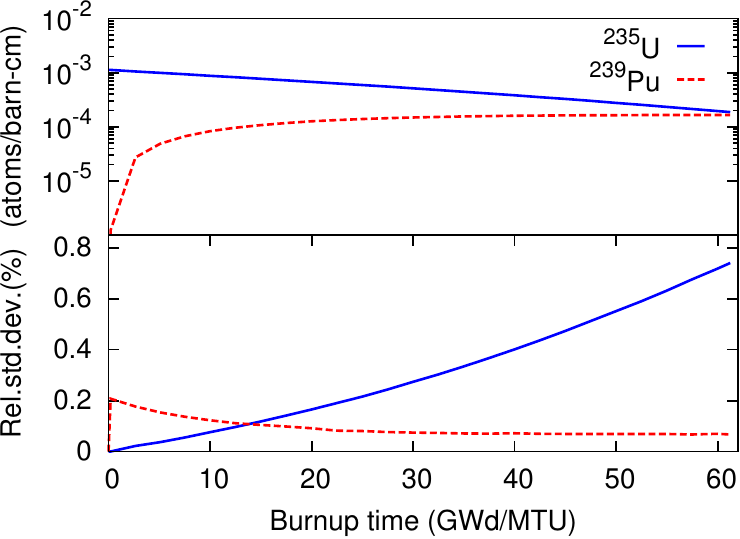}}
  \caption{Uncertainties due to \ura{} cross section uncertainties: the lower panel shows concentrations and uncertainties obtained by a complete NUDUNA analysis; the upper panel shows a comparison of \keff{} uncertainties for a complete analysis and for an analysis that neglects isotopic concentration uncertainties.}
  \label{fig:u235_full_nuduna_vs_tsunami}
\end{figure}
\begin{figure}[htb]
  \centering
  \subfloat[\keff{} uncertainty]{\includegraphics[]{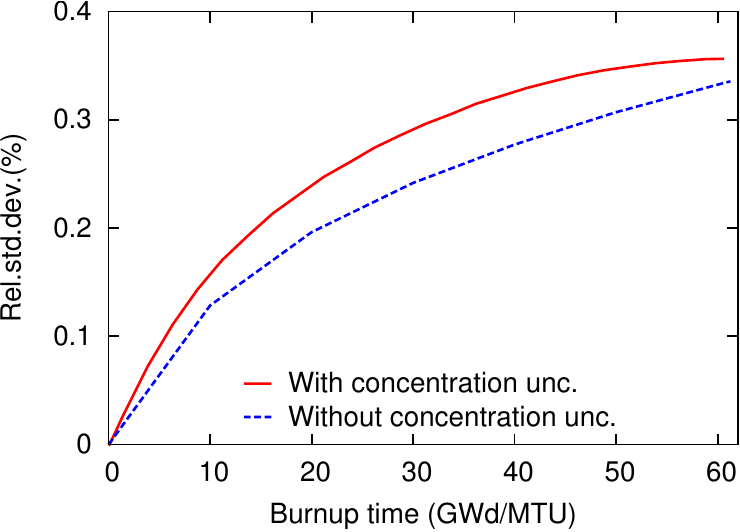}}\\
  \subfloat[Concentrations and their uncertainties]{\includegraphics[]{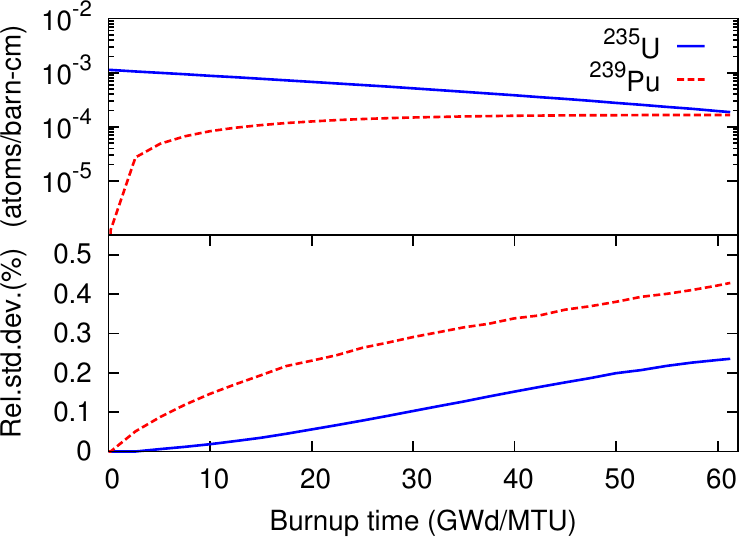}}
  \caption{Same as Fig.~\ref{fig:u235_full_nuduna_vs_tsunami} but due to \plut{} cross section uncertainties.}
  \label{fig:pu239_full_nuduna_vs_tsunami}
\end{figure}
Fig.~\ref{fig:u235_full_nuduna_vs_tsunami} and \ref{fig:pu239_full_nuduna_vs_tsunami} show uncertainties in \keff{} and in the 
isotopic concentrations of \ura{} and \plut{} induced by \ura{} and \plut{} cross section uncertainties, respectively. The lower panels show the isotopic concentrations and their uncertainties as obtained with the full NUDUNA analyzes. The upper panels compare the uncertainty estimates obtained with and without including isotopic concentration uncertainties induced by the depletion step. As can be seen, neglecting the uncertainties of isotopic concentrations leads to a considerable underestimation of the overall uncertainty. 

When propagating $^{235}$U uncertainties, the impact of isotopic concentration uncertainties becomes relevant above 10 GWd/MTU, and increases with increasing $^{235}$U concentration uncertainty. When propagating $^{239}$Pu uncertainties, the omission of isotopic concentration uncertainties shows an effect
already at the very beginning and reaches a maximum between 20 and 50 GWd/MTU.

The lower panels of Fig.~\ref{fig:u235_full_nuduna_vs_tsunami} and
\ref{fig:pu239_full_nuduna_vs_tsunami} also show that $^{235}$U data uncertainties induce isotopic concentration uncertainties on $^{239}$Pu, and vice versa. The reason is that a change in the cross section of one isotope induces changes in neutron flux and spectrum, which modifies the reaction rates of the other isotope whose nuclear data are not modified.

\subsubsection{Impact of  fission neutron multiplicities}
\label{sec:neglecting_n_unc_nubar}
\begin{figure}[tb]
  \centering
  \subfloat[Due to $^{235}$U uncertainties]{\includegraphics[]{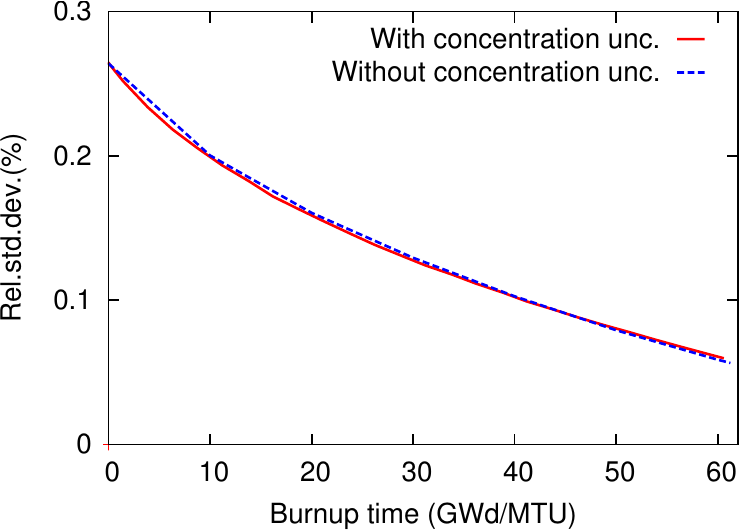}}\\
  \subfloat[Due to $^{239}$Pu uncertainties]{\includegraphics[]{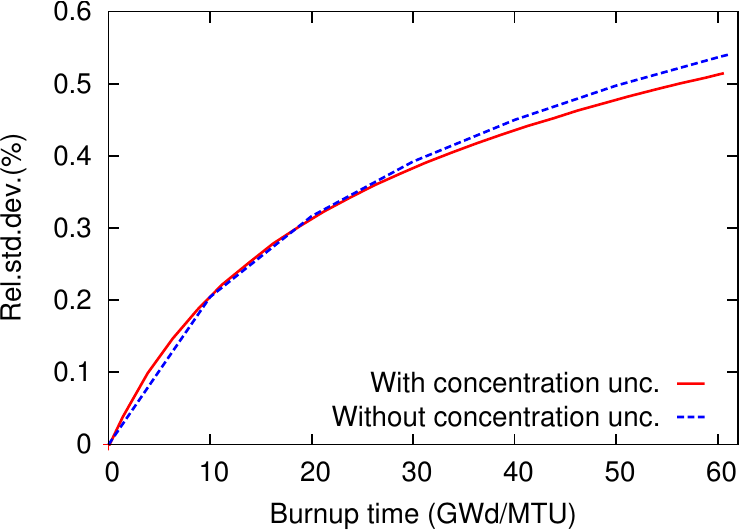}}
  \caption{Uncertainties in \keff{} induced by fission neutron multiplicity (\nubar{}) uncertainties
	    with and without consideration of concentration uncertainties.}
  \label{fig:full_nuduna_vs_tsunami_full_keff}
\end{figure}

Fig.~\ref{fig:full_nuduna_vs_tsunami_full_keff} presents uncertainties induced by fission neutron multiplicity (\nubar{}) uncertainties. Each panel shows a curve obtained by propagating all uncertainties and a curve obtained by neglecting isotopic number density uncertainties.

One observes a very good agreement of the two approximations for \keff{} uncertainties. In fact, it shows that \nubar{} uncertainties have no impact on the depletion step, and also the implicit impact of \nubar{} uncertainties on isotopic compositions via the induced flux uncertainty is negligible.

\subsection{Neglecting neutron flux and spectrum uncertainties}
The last section discussed the impact of neglecting the uncertainties on isotopic concentrations. Now we are going to neglect flux uncertainties. 

Again, NUDUNA is applied to provide the full uncertainty, and the Hybrid Method (HM)~\citep{hybrid_method,cj_andes_nd2013} is used to 
propagate nuclear data uncertainties only in the depletion step. HM is based on Monte Carlo sampling of nuclear data uncertainties, and for each random draw a complete depletion calculation is performed. However, the flux input is kept at its nominal value, and so no additional transport calculations have to be carried out {which implies savings in computing time}. Consequently, neutron flux and spectrum uncertainties are not taken into account.

\begin{figure}[tb]
  \centering
  \subfloat[\ura{} concentration uncertainty]{\includegraphics[]{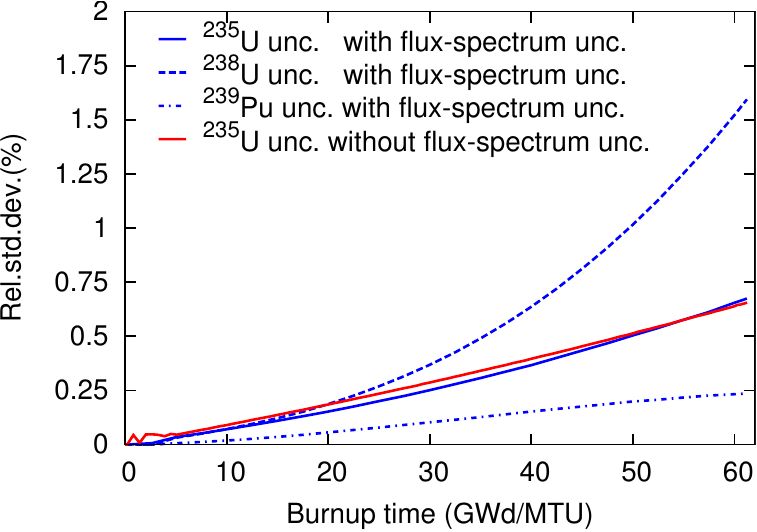}}\\
  \subfloat[$^{236}$U concentration uncertainty]{\includegraphics[]{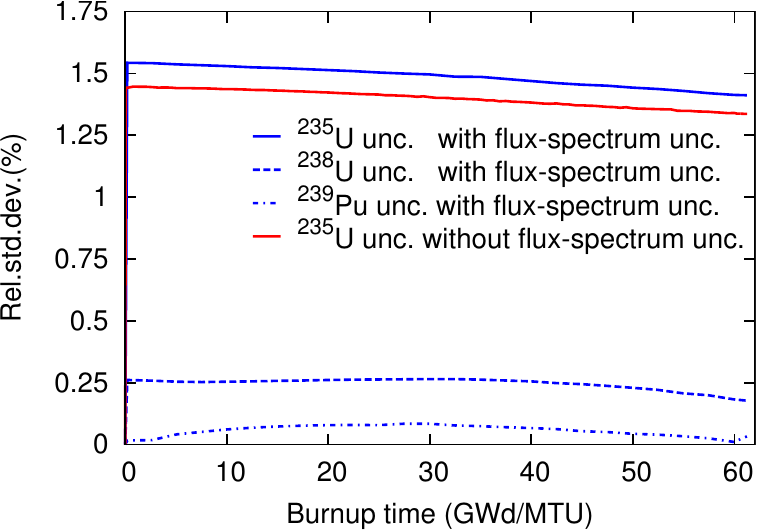}}
  \caption{Uncertainties of \ura{} and $^{236}$U  isotopic concentrations due to $^{235}$U, $^{238}$U, and $^{239}$Pu nuclear data uncertainties obtained with NUDUNA and the Hybrid Method.}
  \label{fig:u235_acab_vs_scale_unc_iso}
\end{figure}

Neglecting neutron flux and spectrum uncertainties implies that the concentration uncertainty of a given isotope is only influenced by its own cross section uncertainty and by those of isotopes that are part of a transmutation chain that results in the given isotope. Fig.~\ref{fig:u235_acab_vs_scale_unc_iso} presents the dominant contributions to the $^{235}$U and $^{236}$U concentration uncertainties.  Indeed, the $^{235}$U concentration addressed in the upper panel is not affected by $^{238}$U or $^{239}$Pu data uncertainties within the HM framework. Propagating also flux uncertainties, as NUDUNA does, leads to sizable contributions of $^{238}$U and $^{239}$Pu data uncertainties to the $^{235}$U concentration uncertainty, as shown by the dashed and dashed-dotted curves in the upper panel of Fig.~\ref{fig:u235_acab_vs_scale_unc_iso}. So the HM method is not capable to predict the $^{235}$U concentration uncertainty since it provides uncertainty estimations much lower than actual ones. The combined effect of 
propagating at the same time the uncertainties in \ura{}, $^{238}$U and \plut{} cross sections has been also addressed, showing that the total uncertainty on the \ura{} concentration is a sum of contributions with no counteracting effects.  
The lower panel of Fig.~\ref{fig:u235_acab_vs_scale_unc_iso} shows the $^{236}$U concentration uncertainty. This isotopic concentration depends via the $^{235}$U(n,$\gamma$)$^{236}$U reaction directly on the $^{235}$U cross sections, and HM yields a good result for the contribution of $^{235}$U data uncertainties to the $^{236}$U concentration uncertainty. Since there are no other  isotopes that sizably impact the $^{236}$U concentration uncertainty, it also gives a  good result for the total uncertainty of the $^{236}$U concentration. 

HM at present does not consider in its random sampling the fact that the reactor power is fixed. Constraining the flux level to this fixed power after random sampling of the cross sections will induce a variation on the flux level. This could possibly lead to an improved HM uncertainty estimate, and future studies should address this topic. 

To summarize, neglecting neutron flux and spectrum uncertainties may lead to considerable underestimation of the overall concentration uncertainty. However, there are cases where such an approximation yields good results. Given the gains in computing time, future studies might also address applicability criteria of HM such that HM could at least be used to study a limited set of isotopes.


\section{Impact of covariance data input}

The previous section was solely based on \scale{} covariance data input in order to compare NUDUNA results to SCALE TSUNAMI and Hybrid Method results. Next, the impact of different covariance data input  on the obtained uncertainty estimates is addressed. As outlined in Sec.~\ref{sec:why_scale60_endfb71}, \efb{} provides the most modern covariance evaluation which also includes all fuel isotopes. The \scale{} and \efb{} libraries show major differences for the important \ura{} and \plut{} fuel isotopes, and the following paragraphs address these isotopes as examples to demonstrate the need for consistent covariance data input.

\subsection{\efb{} and \scale{} uncertainties for \ura{} and \plut{}}
\label{sec:diff_scale60_endfb71}
\begin{figure*}[p]
  \centering
  \subfloat[\efb{} total \ura{} \nubar{}]{\resizebox{0.68\columnwidth}{!}{\rotatebox{-90}{\includegraphics[page=1]{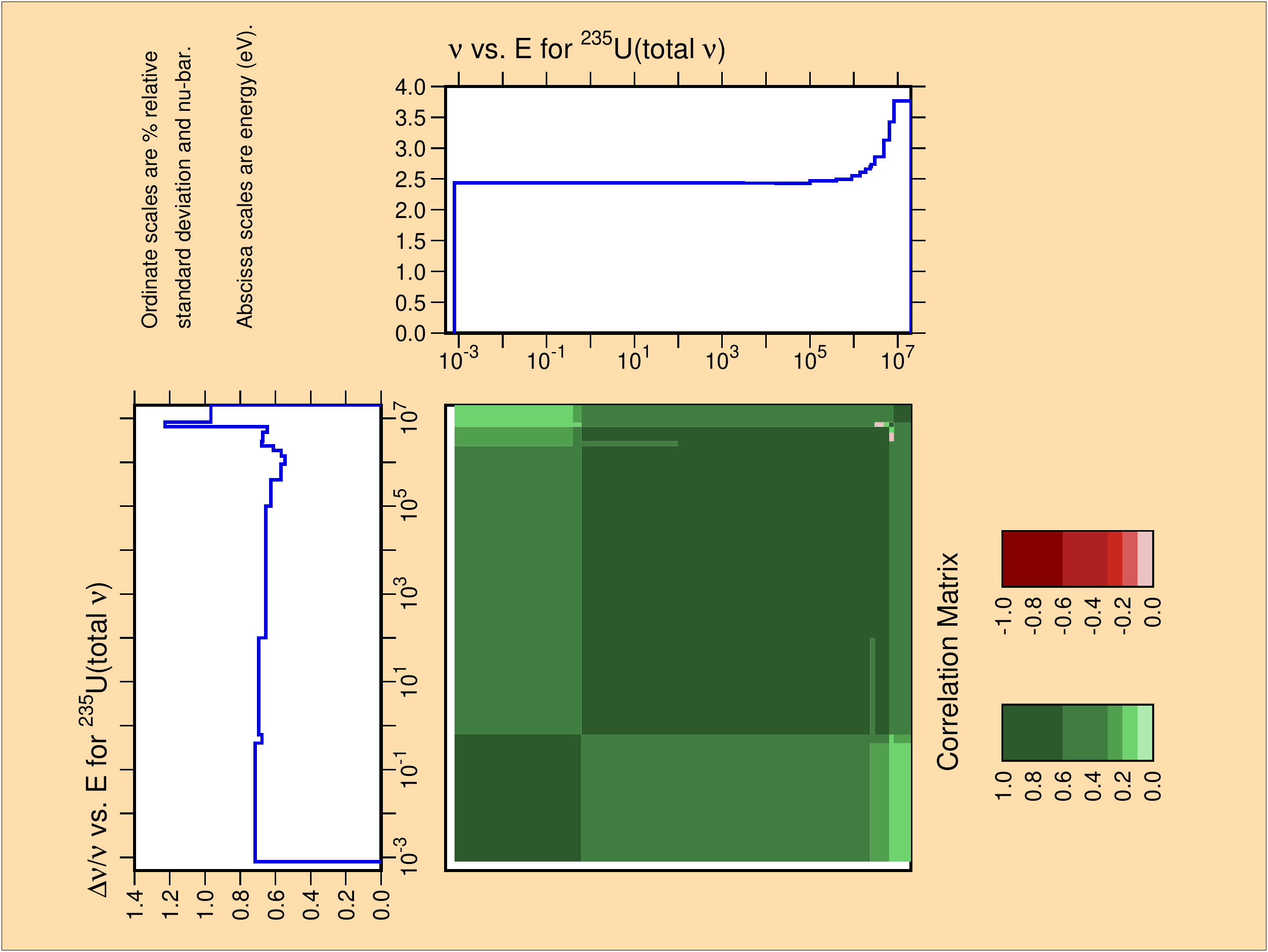}}}}
  \subfloat[\efb{} prompt \ura{} \nubar{}]{\resizebox{0.68\columnwidth}{!}{\rotatebox{-90}{\includegraphics[page=2]{u235-endfb71-nubar-covariances-crop.pdf}}}}
  \subfloat[\scale{} \ura{} \nubar{}]{\resizebox{0.68\columnwidth}{!}{\rotatebox{-90}{\includegraphics[]{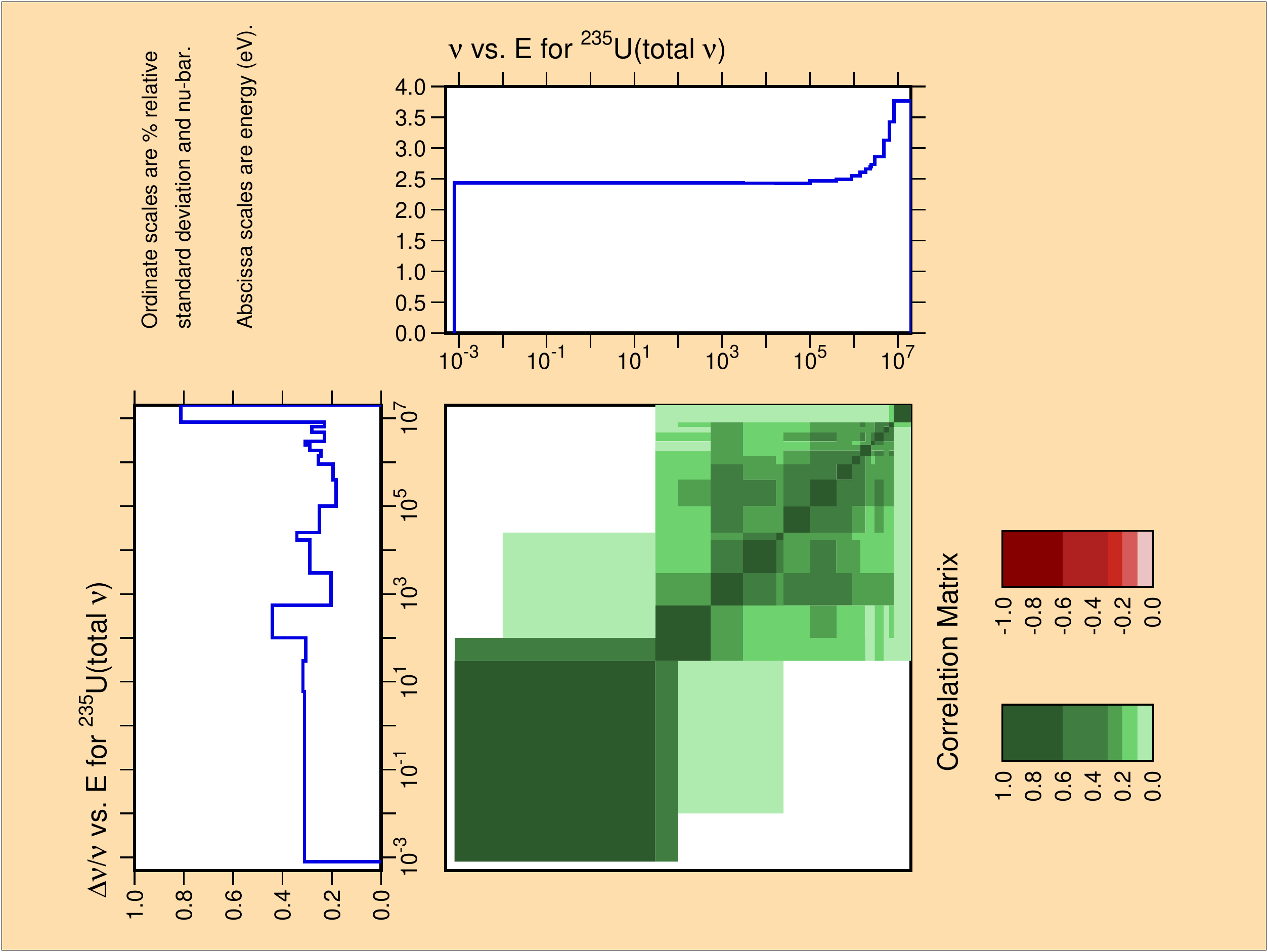}}}}\\

\vspace{0.5cm}

\begin{minipage}{0.98\columnwidth}
\raggedleft
  \subfloat[\efb{} total \plut{} \nubar{}]{\resizebox{0.68\columnwidth}{!}{\rotatebox{0}{\includegraphics[]{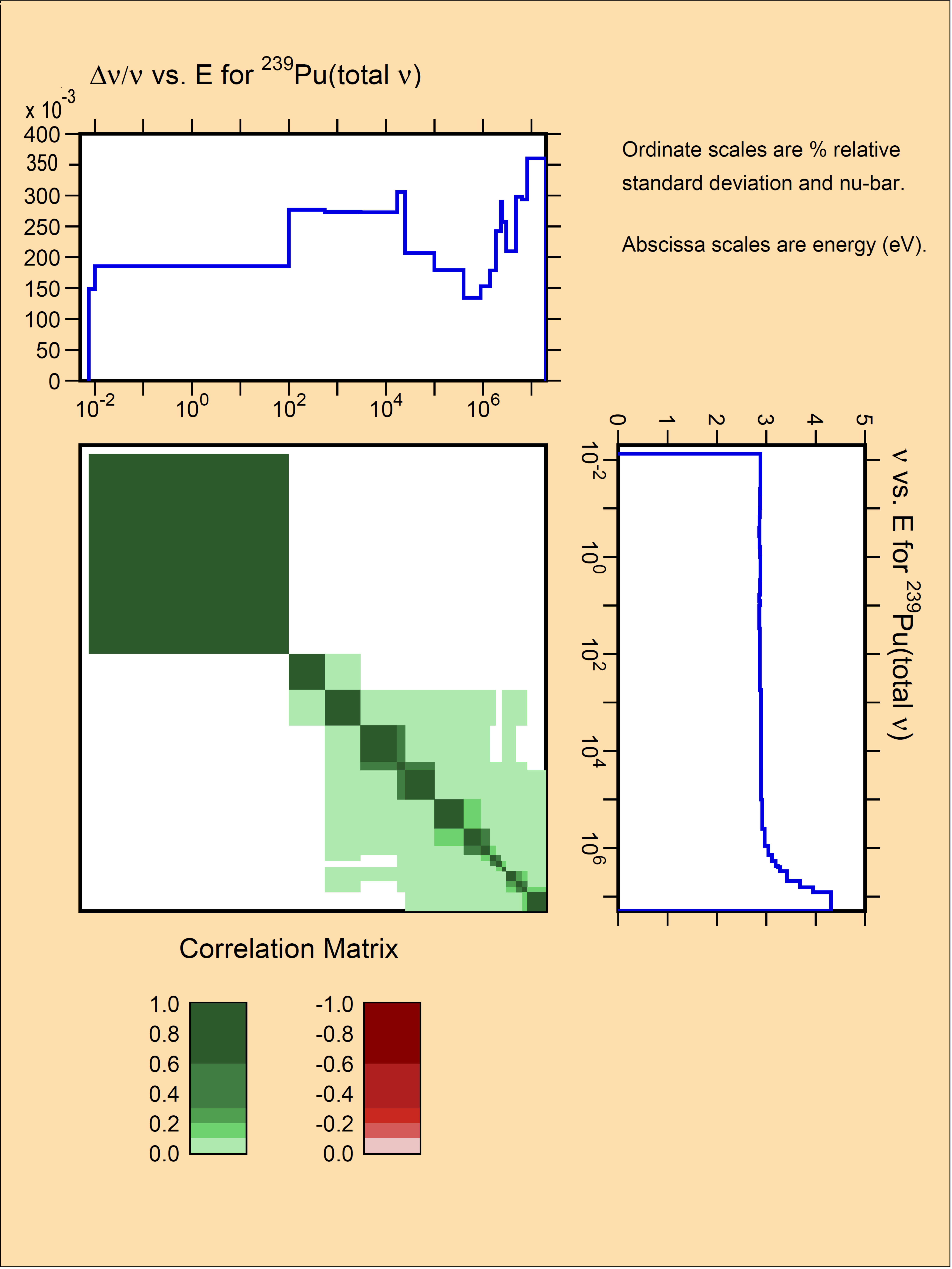}}}}
\end{minipage}\qquad
\begin{minipage}{0.98\columnwidth}
\raggedright
 \subfloat[\scale{} \plut{} \nubar{}]{\resizebox{0.68\columnwidth}{!}{\rotatebox{-90}{\includegraphics[page=1]{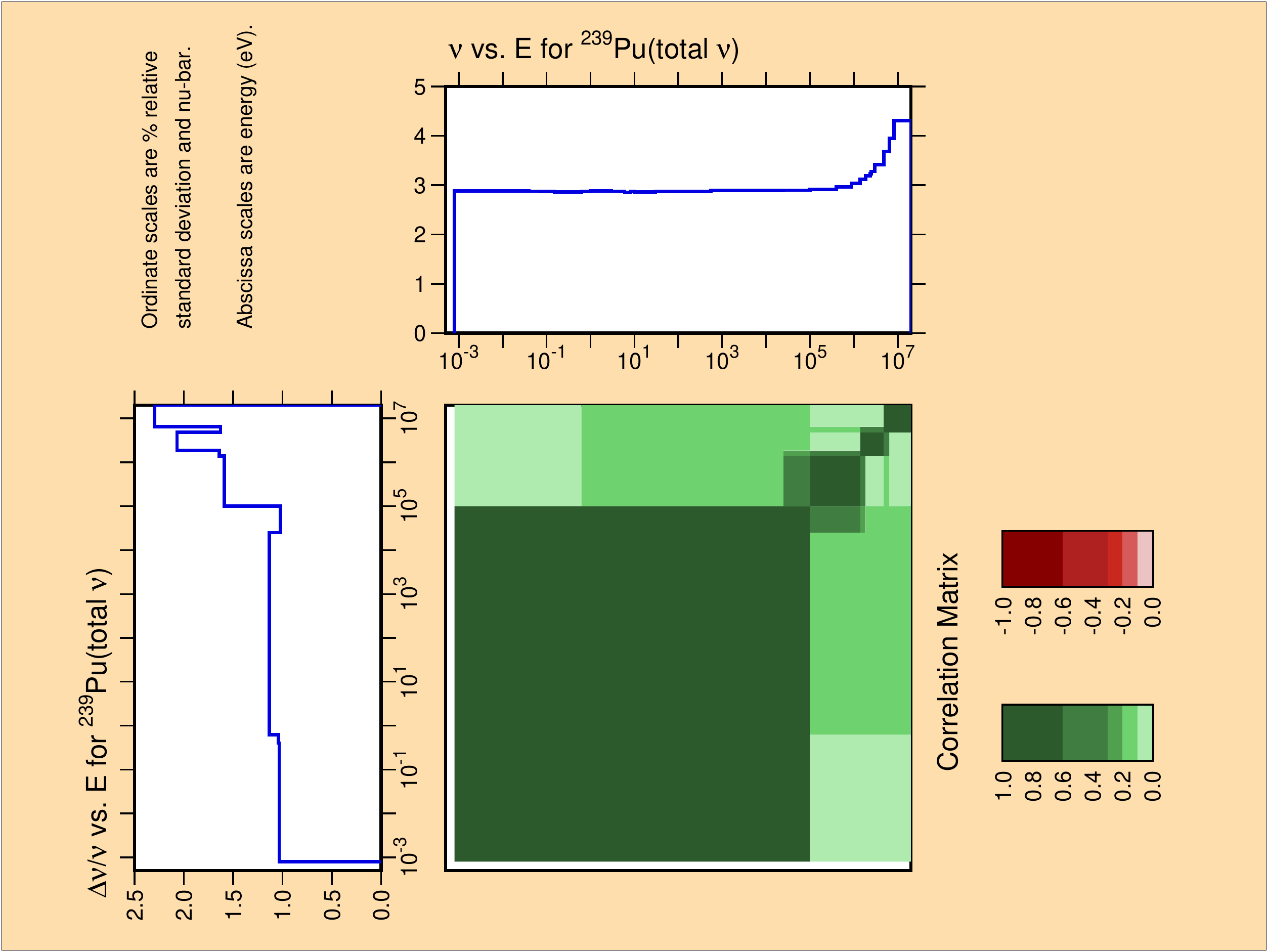}}}}
\end{minipage}

  \caption{Covariance data for \ura{} (upper panels) and \plut{} (lower panels) fission neutron multiplicities from \efb{} and \scale{}, presented in the SCALE 44-group structure.}
  \label{fig:u235_nubar_cov_endfb71}
\end{figure*}

The upper panels of Fig.~\ref{fig:u235_nubar_cov_endfb71} show $^{235}$U \nubar{}-uncertainties as provided by \efb{} and \scale{}. \efb{} gives covariance data for both prompt (\nubar{p}) and total (\nubar{t}) neutron multiplicities. However, the two covariance matrices are not consistent: the difference between the covariance matrices is too large to be explained by the delayed neutron multiplicity \nubar{d} whose contribution to the total multiplicity is only approx.~$0.64\%$.  Hence, the total and prompt covariance matrices should be almost identical. There are also major differences between \scale{} and \efb{} covariance data for \nubar{}. With regards to \ura{} cross section uncertainties, only small differences are found among \efb{} and \scale{}.

The lower panels of Fig.~\ref{fig:u235_nubar_cov_endfb71} show the \nubar{} uncertainties for $^{239}$Pu. For \efb{}, the covariance matrices for the total and prompt multiplicities are identical and, consequently, only the one for the total multiplicity is presented. Again, very large differences are observed between \efb{} and \scale{}. Additionally, the \efb{} and \scale{} cross section uncertainties differ
for (n,fission) $[$E$>$1eV$]$, (n,$\gamma$) $[$1-100eV$]$ and the cross-correlation between (n,fission) and (n,$\gamma$).

\subsection{Propagating \efb{} and \scale{} uncertainties for \ura{} and \plut{}}
\begin{figure}[tb]
  \centering
  \subfloat[Due to $^{235}$U uncertainties\label{fig:nubar_keff_unc_1}]{\includegraphics[]{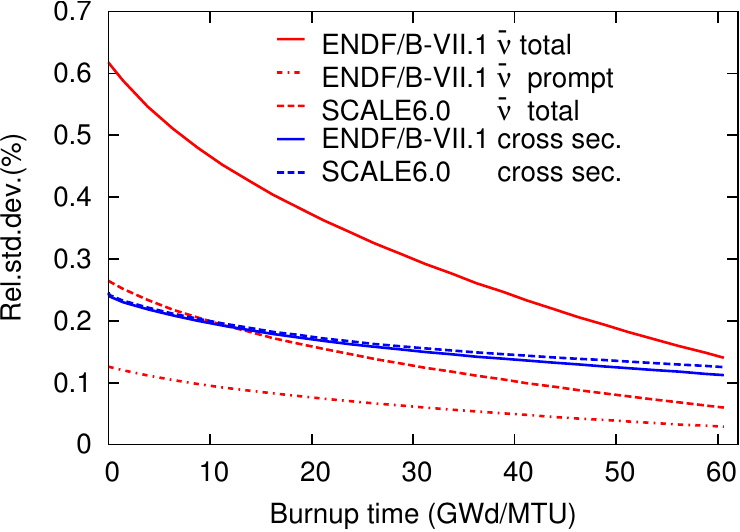}}\\
  \subfloat[Due to $^{239}$Pu uncertainties\label{fig:nubar_keff_unc_2}]{\includegraphics[]{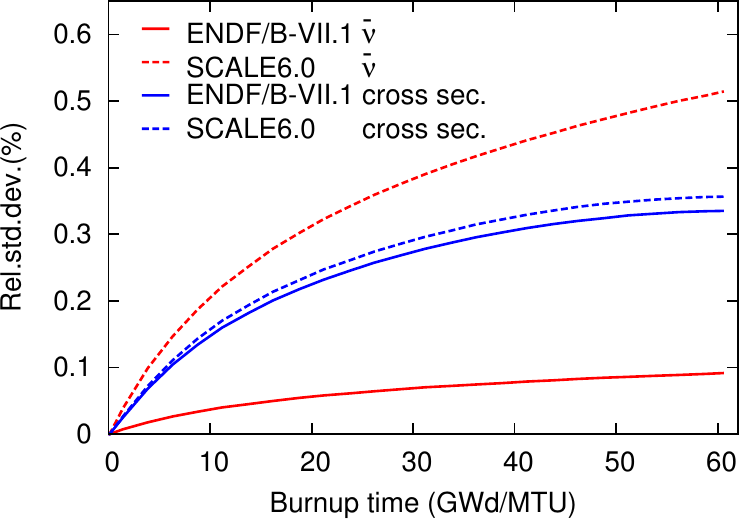}}
  \caption{Contributions of \nubar{} and cross section uncertainties to the \keff{} uncertainty as a function of burn-up, obtained with NUDUNA.}
  \label{fig:nubar_keff_unc}
\end{figure}

Fig.~\ref{fig:nubar_keff_unc} presents the \keff{} uncertainties induced by \ura{} and \plut{} cross section and \nubar{} uncertainties as predicted by NUDUNA. The SCALE uncertainties have been obtained by creating an \ef{} formatted file that represents a fusion of ENDF/B-VII.1 nominal information and SCALE uncertainty information. 

Two major conclusions can be drawn from these results. First, the uncertainty estimates for \keff{} strongly depend on the choice of the input source: the \nubar{} contributions based on \efb{} uncertainties are very different from those based on \scale{} uncertainties. The second conclusion is related to the fact that in \efb{} the uncertainties for total and prompt \nubar{} for $^{235}$U  are inconsistent (cf.~Sec.~\ref{sec:diff_scale60_endfb71}), so that one has to choose between sampling the \nubar{} data based on the total or on the prompt uncertainties. As shown by a comparison of the red solid and the red dashed-dotted curve in Fig.~\ref{fig:nubar_keff_unc_1}, this arbitrary choice induces a large ambiguity in the predicted contribution of the $^{235}$U \nubar{} uncertainty to the \keff{} uncertainty. 

Fig.~\ref{fig:nubar_keff_unc} also shows results for propagating  cross section uncertainties, i.e.~for propagating File~33 (MF33) covariance information. For \ura{}, the \efb{} and \scale{} results are very similar, as could be expected from their similar covariance information. However, the results are also very similar for $^{239}$Pu, in spite of the fact that there are significant differences in the $^{239}$Pu covariances. Here it seems that differences between individual cross sections counteract each other.

To summarize, depending on the choice of uncertainty data input, \efb{} or \scale{}, very different UQ results are obtained. Even the most important uncertainty contributors may change. Additionally, the  inconsistency in \nubar{} uncertainties for $^{235}$U in \efb{} has to be addressed.


\section{Full UQ analysis based on \efb{}}
In the following, an UQ analysis with focus on nuclear data is performed, which includes almost all contributions to nuclear data uncertainty. The contribution from $S(\alpha,\beta)$ uncertainties is, however, not included, since these uncertainties are not yet fully quantified and not yet provided within evaluated nuclear data libraries. Currently, methodologies are being developed for estimating them~\citep{Sab_nrg,Sab_usa}, and there are also first impact estimates available, e.g., by ~\cite{oscar_net_kns_2014} for PWR reactor cores. 

In order to perform the UQ analysis, the incident neutron data and the decay data were sampled according to \efb{} uncertainty data. The decay data sampling considered all isotopes included in the ORIGEN input, while the incident neutron data sampling was limited to the most important isotopes presented in Table~\ref{tab:list_iso}, and data for the minor isotopes were not sampled but taken from the nominal ENDF/B-V library of \scale. 

\begin{figure}[tb]
  \centering
  \includegraphics[]{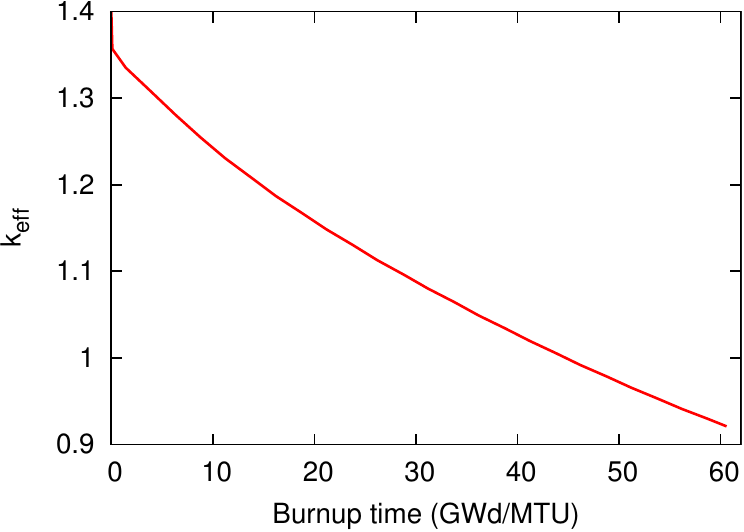}
  \caption{Evolution of \keff{} as a function of burn-up. The neutron data are taken from \efb{} for nuclides given in Table~\ref{tab:list_iso}, and from the \scale{} ENDF/B-V multi-group cross section library for all others. Decay data are taken from \efb{} for all isotopes.}
  \label{fig:keff_ref}
\end{figure}

A reference calculation is presented in Fig.~\ref{fig:keff_ref}, which is based on nominal \efb{} decay data, nominal \efb{} neutron data for the isotopes in Table~\ref{tab:list_iso} and nominal ENDF/B-V neutron data for the other isotopes. The evolution of the neutron multiplication factor \keff{} presented in Fig.~\ref{fig:keff_ref} shows the typical PWR pin-cell behavior.

\begin{table}[tb]
  \centering
  \caption{Isotopes whose incident neutron data and uncertainties are included in the full UQ analysis.}
  \label{tab:list_iso}
  \footnotesize
  \begin{tabular}{cc}
    \hline
    \hline
Light isotopes &  $^1$H, $^{16}$O, $^{99}$Tc, $^{133}$Cs \\
     {and}          &  $^{143}$Nd, $^{135}$Xe, $^{149}$Sm, $^{151}$Sm \\
 {fission products} &  $^{151}$Eu, $^{155}$Eu, $^{155}$Gd, $^{157}$Gd \\
\hline
Heavy isotopes &  $^{234}$U, $^{235}$U, $^{236}$U, $^{238}$U \\
               &  $^{239}$Pu, $^{240}$Pu, $^{241}$Pu \\
    \hline
    \hline
  \end{tabular}
\end{table}

The following paragraphs discuss the impacts of incident neutron and decay data uncertainties separately in order to illustrate their individual contributions.

\subsection{Impact of \efb{} incident neutron uncertainties}
\begin{figure}[tb]
  \centering
  \includegraphics[]{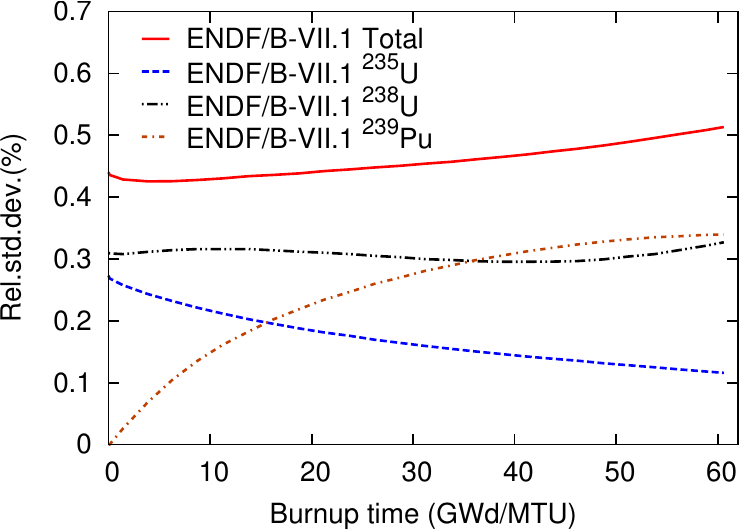}
  \caption{\keff{} uncertainty as a function of burn-up when propagating 
	   incident neutron data uncertainties according to \efb{} covariance data.}
  \label{fig:endfb71_keff_unc}
\end{figure}

Fig.~\ref{fig:endfb71_keff_unc} shows the \keff{} uncertainty estimate that has been obtained based on the \efb{} incident neutron data covariances for the list of isotopes given in Table~\ref{tab:list_iso}. This estimate includes uncertainties of \nubar{}, resonance parameters, cross sections, and angular distributions. The \nubar{} values are sampled based on the covariance information for the prompt multiplicities if there is covariance information for both prompt and total multiplicities.

Fig.~\ref{fig:endfb71_keff_unc} also presents the individual contributions due to $^{235}$U, $^{238}$U and $^{239}$Pu uncertainties. The  $^{238}$U data uncertainty causes a prominent almost constant \keff{} uncertainty contribution throughout the whole burn-up period. The total uncertainty at begin of cycle (BOC) amounts to $440$~pcm and decreases towards the end of cycle (EOC) due to a decreasing $^{235}$U uncertainty contribution. Then fission products and Plutonium isotopes become relevant, and the slope of the total \keff{} uncertainty becomes positive and amounts to approximately $510$~pcm at EOC. 

The above estimate does not include uncertainties induced by the fission neutron spectrum $\chi$ {and by  the fission yields since their sampling} has not yet been implemented in NUDUNA. However, we have  estimated the fission neutron spectrum with the aid of TSUNAMI. The contribution of the \ura{} $\chi$ uncertainty to \keff{} decreases from $100$~pcm at BOC to $50$~pcm at EOC, and the \plut{} $\chi$ uncertainty contribution increases from $36$~pcm at 10 GWd/tHM to $150$~pcm at EOC. Adding these contributions quadratically to the pure NUDUNA estimates results in the uncertainty estimates of approximately $450$~pcm at BOC and $530$~pcm at EOC. So the fission neutron spectrum $\chi$ adds an additional contribution, but it is less than $5$\% of the total uncertainty. {Based on the latest estimates for fission yield uncertainties~\cite{luca_fy_paper}, it has been shown in \cite{oscar_jeff_1566} that this contribution yields very small uncertainty contributions to \keff{} (circa 50~pcm throughout the 
complete burn-up). However, they have impact on isotopic composition uncertainties~\cite{oscar_jeff_1566} and their sampling should be subject of a future NUDUNA enhancement.}

In Table~\ref{tab:unc_iso_comp}, results for uncertainties of isotopic compositions induced by incident neutron data uncertainties are presented. As can be seen, uncertainties for uranic and transuranic isotopes do not exceed 4\% at EOC, and the ones for fission products stay below 8\% for almost all isotopes. Two exceptions are $^{155}$Eu and $^{155}$Gd for which the uncertainties amount to 36\% and 27\%, respectively. These large uncertainties are caused by large $^{155}$Eu (n,$\gamma$) cross section uncertainties, which also affects the $^{155}$Gd isotope being a decay product of $^{155}$Eu. Additionally, the (n,$\gamma$) reaction in $^{155}$Gd has uncertainties of approx.~$10$\%.

\begin{table*}[tb]
  \centering
  \caption{Isotopic composition uncertainties as a function of burn-up when propagating 
	   incident neutron data uncertainties according to \efb{} covariance data (only those isotopes are shown for which the uncertainty exceeds 1\%).}
  \footnotesize
  \label{tab:unc_iso_comp}
  \begin{tabular}{*{10}c}
    \hline
    \hline
	burn-up$\rightarrow$&     \multicolumn{1}{c}{0 GWd/MTU}                  & \multicolumn{2}{c}{10 GWd/MTU} & \multicolumn{2}{c}{30 GWd/MTU} & \multicolumn{2}{c}{50 GWd/MTU} & \multicolumn{2}{c}{60 GWd/MTU} \\
Isotope & (at./barn-cm) & (at./barn-cm) & unc.(\%) & (at./barn-cm) & unc.(\%) & (at./barn-cm) & unc.(\%) & (at./barn-cm) & unc.(\%) \\
    \hline
$^{134}$Cs & 0.000E+00 & 4.442E-07 & 4.71 & 3.362E-06 & 4.38 & 7.824E-06 & 4.06 & 1.034E-05 & 3.89 \\
$^{143}$Nd & 0.000E+00 & 1.228E-05 & 0.29 & 3.237E-05 & 0.91 & 4.499E-05 & 1.60 & 4.879E-05 & 1.97 \\
$^{149}$Sm & 0.000E+00 & 1.176E-07 & 5.07 & 1.267E-07 & 5.24 & 1.181E-07 & 5.46 & 1.127E-07 & 5.58 \\
$^{151}$Sm & 0.000E+00 & 4.014E-07 & 5.50 & 5.858E-07 & 6.64 & 6.966E-07 & 6.90 & 7.381E-07 & 6.98 \\
$^{151}$Eu & 0.000E+00 & 5.963E-10 & 4.85 & 9.179E-10 & 6.63 & 9.634E-10 & 7.06 & 9.538E-10 & 7.26 \\
$^{155}$Eu & 0.000E+00 & 4.567E-08 & 27.85 & 2.126E-07 & 33.21 & 4.789E-07 & 35.36 & 6.113E-07 & 35.58 \\
$^{155}$Gd & 0.000E+00 & 5.243E-10 & 27.26 & 2.456E-09 & 29.89 & 5.268E-09 & 28.29 & 6.571E-09 & 26.74 \\
$^{156}$Gd & 0.000E+00 & 1.585E-07 & 7.17 & 1.352E-06 & 4.82 & 4.752E-06 & 3.28 & 7.540E-06 & 2.64 \\
\hline
$^{234}$U & 1.166E-05 & 1.041E-05 & 0.31 & 8.153E-06 & 0.98 & 6.231E-06 & 1.74 & 5.406E-06 & 2.15 \\
$^{235}$U & 1.126E-03 & 8.764E-04 & 0.11 & 5.137E-04 & 0.48 & 2.770E-04 & 1.22 & 1.958E-04 & 1.78 \\
$^{236}$U & 0.000E+00 & 4.592E-05 & 1.50 & 1.076E-04 & 1.44 & 1.402E-04 & 1.40 & 1.480E-04 & 1.39 \\
$^{237}$Np & 0.000E+00 & 1.618E-06 & 4.10 & 8.397E-06 & 3.34 & 1.622E-05 & 3.11 & 1.970E-05 & 3.03 \\
$^{238}$Pu & 0.000E+00 & 1.218E-07 & 4.60 & 2.020E-06 & 3.41 & 6.985E-06 & 3.01 & 1.039E-05 & 2.89 \\
$^{239}$Pu & 0.000E+00 & 8.256E-05 & 1.15 & 1.485E-04 & 1.43 & 1.639E-04 & 1.81 & 1.648E-04 & 2.00 \\
$^{240}$Pu & 0.000E+00 & 9.675E-06 & 1.57 & 4.005E-05 & 1.78 & 6.514E-05 & 2.01 & 7.425E-05 & 2.12 \\
$^{241}$Pu & 0.000E+00 & 3.802E-06 & 1.49 & 2.570E-05 & 1.43 & 4.317E-05 & 1.72 & 4.851E-05 & 1.91 \\
$^{242}$Pu & 0.000E+00 & 2.162E-07 & 2.72 & 5.195E-06 & 2.46 & 1.644E-05 & 2.41 & 2.339E-05 & 2.43 \\
$^{243}$Am & 0.000E+00 & 1.066E-08 & 2.95 & 8.629E-07 & 2.59 & 4.334E-06 & 2.36 & 7.028E-06 & 2.26 \\
$^{244}$Cm & 0.000E+00 & 6.472E-10 & 3.21 & 1.900E-07 & 2.92 & 1.802E-06 & 2.73 & 3.678E-06 & 2.67 \\
    \hline
    \hline
  \end{tabular}
\end{table*}

\subsection{Impact of \efb{} decay data uncertainties}
Now we are going to study the uncertainties induced by \efb{} decay data uncertainties, i.e.~uncertainties of
half-lives and branching ratios.

Any data without uncertainty specification in \efb{} is assigned a 100\% uncertainty in the NUDUNA random sampling procedure. Even with this conservative assumption, the resulting \keff{} uncertainty  is completely negligible and does not exceed 6~pcm during the whole burn-up period.

The impact of decay data uncertainties on isotopic compositions is also limited, and only for three isotopes the induced uncertainty exceeds
0.3\%: $^{133}$Cs, $^{134}$Cs and $^{151}$Eu. The large uncertainty of $^{151}$Eu, whose number density uncertainty is almost constant at 9.3\% throughout the burn-up period, has been identified before in \citep{oscar_ans_2011,oscar_icnc_2011,cj_andes_nd2013}, and its large uncertainty comes from the $6.67\%$ uncertainty on the $^{151}$Sm half-life.

\begin{figure}[t]
  \centering
  \includegraphics[]{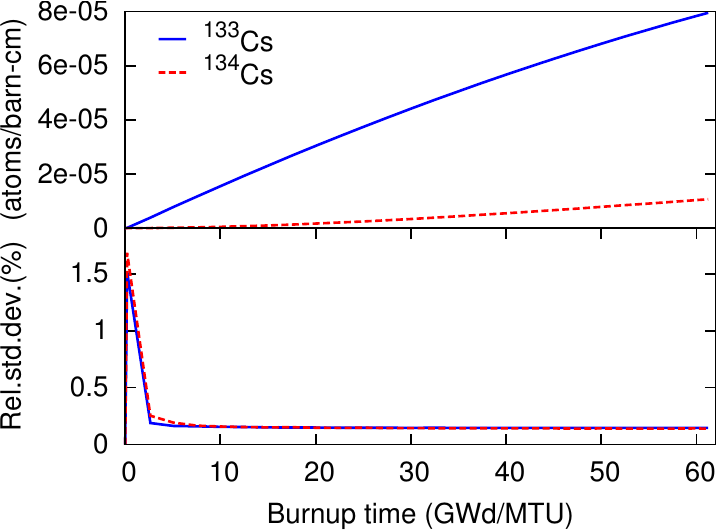}
		\caption{Uncertainties of Cesium isotope concentrations as a function of burn-up when propagating decay data uncertainties.}
  \label{fig:endfb71_iso_unc_dd}
\end{figure}

The $^{133}$Cs and $^{134}$Cs uncertainties are presented in Fig.~\ref{fig:endfb71_iso_unc_dd}. For $^{134}$Cs the decay of its parent $^{134}$Xe plays a minor role since $^{134}$Xe is almost stable (half-life $~5.8\times10^{22}$ years), and it is mainly
produced by neutron capture in $^{133}$Cs. Thus, the $^{134}$Cs uncertainty follows the $^{133}$Cs uncertainties. $^{133}$Cs is a stable isotope and its number density uncertainties are generated by the decay data uncertainties of its parents $^{133}$Xe, $^{133}$I, and $^{133}$Te. These parent isotopes have small half-life values and cause already at BOC quite a large uncertainty of their $^{133}$Cs daughter number density.

Table~\ref{tab:dd_unc_cooling_time} shows the uncertainty estimates for the number densities after 300~years of cooling. They are only sizable for the $^{137}$Cs, $^{151}$Eu, $^{151}$Sm, and $^{241}$Pu isotopes. These results are in agreement with the ones provided by \cite{jsg_nd2013} who also applied a Monte Carlo based method, except for $^{137}$Cs for which we obtain only half of the uncertainty.

\begin{table}[H]
  \centering
  \caption{Uncertainties in isotopic number densities due to \efb{} decay data uncertainties after 300~years of cooling.}
  \label{tab:dd_unc_cooling_time}
  \footnotesize
  \begin{tabular}{ccc}
    \hline
    \hline
  Isotope & (at./barn-cm) &unc.(\%) \\
  \hline
  $^{137}$Cs & 8.739E-08 & 1.86 \\
  $^{151}$Eu & 6.690E-07 & 2.26 \\
  $^{151}$Sm & 7.383E-08 & 21.16 \\
  $^{241}$Pu & 6.246E-10 & 1.05 \\
    \hline
    \hline
  \end{tabular}
\end{table}


\section{Conclusions}
This work presented an uncertainty quantification study for the burn-up of a typical PWR pin-cell, focusing on uncertainties induced by nuclear data uncertainties. Different approximations to uncertainty assessment were studied by evaluating their performance in the burn-up analysis of the UAM \exercise{} pin-cell benchmark.

An analysis of the \keff{} uncertainty for transport calculations at different burn-up points based on first order approximation leads to results that are very similar to the exact results obtained by Monte Carlo sampling when propagating \ura{} and \plut{} uncertainties. Note that in \citep{oscar_stni_2013} and \citep{sabouri_nd2013} first order uncertainty propagation methods are proposed that include both uncertainties of transport and depletion. However, these schemes yield higher \keff{} uncertainty estimates than the exact Monte Carlo schemes.

Neglecting either flux or isotopic number density uncertainties in the burn-up simulation is both problematic and may lead to considerable underestimations of the total uncertainty. Neglecting uncertainties in the isotopic number densities leads to sizable deviations for the \keff{} uncertainty, especially at EOC. Neglecting the flux uncertainties, as proposed by the Hybrid Method, leads to an underestimation of the overall isotopic number density uncertainty. The magnitude of the underestimation is hard to predict, and so it is not clear when such an approximation can be applied in order to perform fast UQ studies for burn-up problems. Future HM developments for defining applicability criteria and for including a reactor power constraint in the random sampling are desirable. Overall, we conclude that accurate uncertainty quantification studies should always fully propagate flux and isotopic number density uncertainties.

Additionally, the impact of different covariance input data was analyzed with NUDUNA. For \ura{}, an inconsistency between total and prompt \nubar{} uncertainties is found for \efb, which leads to ambiguous results for the resulting \keff{} uncertainty. Owing to the large differences between \scale{} and \efb{} \linebreak[4]\nubar{}-uncertainties, the \keff{} uncertainty study based on \linebreak[4]\scale{} yields much higher uncertainties than the one based on \efb{}. 

Finally, a complete  uncertainty quantification was performed with NUDUNA for the UAM benchmark based on \efb{} covariances. The \keff{} uncertainty amounts to approx.~$450$~pcm at BOC and approx.~$530$~pcm at EOC. The most important contributions stem from $^{235}$U, $^{238}$U, $^{239}$Pu data uncertainties: the $^{235}$U contribution decreases and the $^{239}$Pu contribution increases as burn-up increases, and the contribution of $^{238}$U is almost constant until EOC. The impact of decay data on the \keff {}uncertainty is negligible.

The incident neutron data uncertainties lead to uncertainties in isotopic concentrations. The number density uncertainties typically do not exceed 4\% for heavy isotopes and 8\% for fission products. Only for $^{155}$Eu and $^{155}$Gd higher values of up to 35\% are obtained. Decay data uncertainties induce additional sizable uncertainties for the $^{137}$Cs, $^{151}$Eu, $^{151}$Sm, and $^{241}$Pu isotopes, and the results for the uncertainties 300 years after shutdown are in good agreement with the ones obtained by \cite{jsg_nd2013}.

Monte Carlo schemes are up to now the only ones capable of including both uncertainties of neutron transport and depletion calculations. The presented Monte-Carlo scheme may also be extended to uncertainty assessments on a fuel assembly or reactor level.

Note that this work represents an exemplary UQ study based on currently available covariance data. The study should be repeated once updated covariance data are provided or new ENDF/B or SCALE libraries are released.


\section*{Acknowledgements}
This work has been partially supported by ``Ministerio de Educaci\'on (Ministry of Education)''
 of Spain  with the FPU Program for teaching and researching formation
(Programa de Formaci\'on de Profesorado Universitario) under grant AP2009-1801 for the first author,
 and by CSN, project on uncertainties P120530507, and was conducted in the framework of the  AREVA~GmbH R\&D project "`Uncertainty Analysis and Uncertainty Propagation in Nuclear Design Systems"'. Consejo Social UPM 2014 provided funding for the first author's stay at AREVA GmbH between 01/2014-04/2014. We thank M.~Lamm from AREVA~GmbH for his support in R\&D project management.


\bibliographystyle{elsarticle-harv}
\bibliography{bibtex}

\begin{thebibliography}{46}
\expandafter\ifx\csname natexlab\endcsname\relax\def\natexlab#1{#1}\fi
\expandafter\ifx\csname url\endcsname\relax
  \def\url#1{\texttt{#1}}\fi
\expandafter\ifx\csname urlprefix\endcsname\relax\def\urlprefix{URL }\fi

\bibitem[{Buss et~al.(2011)Buss, Hoefer, and Neuber}]{nuduna}
Buss, O., Hoefer, A., Neuber, J.~C., 2011. {NUDUNA - Nuclear Data Uncertainty
  Analysis}. In: Proc. International Conference on Nuclear Criticality (ICNC
  2011), Edinburgh, Scotland.

\bibitem[{Cabellos(2013)}]{oscar_stni_2013}
Cabellos, O., 2013. {Presentation and Discussion of the UAM Exercise I-1b:
  Pin-Cell Burn-Up Benchmark with the Hybrid Method}. Science and Technology of
  Nuclear Installations 2013, 12, article ID 790206.

\bibitem[{Cabellos et~al.(2014{\natexlab{a}})Cabellos, Castro, Ahnert, and
  Holgado}]{oscar_net_kns_2014}
Cabellos, O., Castro, E., Ahnert, C., Holgado, C., june 2014{\natexlab{a}}.
  {Propagation of Nuclear Data Uncertainties for PWR core analysis}. Nuclear
  Engineering and Technology 46~(3).

\bibitem[{Cabellos et~al.(2011{\natexlab{a}})Cabellos, Mart\'inez, and
  D\'iez}]{oscar_ans_2011}
Cabellos, O., Mart\'inez, J.~S., D\'iez, C.~J., 2011{\natexlab{a}}. {Impact of
  nuclear data uncertainties in the Phase-1B Benchmark}. Transactions- American
  Nuclear Society 104, 369 -- 370.

\bibitem[{Cabellos et~al.(2011{\natexlab{b}})Cabellos, Mart\'nez, and
  D\'iez}]{oscar_icnc_2011}
Cabellos, O., Mart\'nez, J.~S., D\'iez, C.~J., Sept. 19-23 2011{\natexlab{b}}.
  {Isotopic uncertainty assessment due to nuclear data uncertainties in
  high-burnup samples}. In: Int. Conf. on Nuclear Criticality ({ICNC} 2011).
  Edimburgo, Escocia.

\bibitem[{Cabellos et~al.(2014{\natexlab{b}})Cabellos, Piedra, and
  D\'iez}]{oscar_jeff_1566}
Cabellos, O., Piedra, D., D\'iez, C.~J., April 2014{\natexlab{b}}. {Impact of
  the fission yield covariance data in burn-up calculations}. Tech. Rep.
  {JEF/DOC-1566}, {OECD/NEA Data Bank}.

\bibitem[{Cacuci(2003)}]{cacuci}
Cacuci, D.~G., 2003. {Sensitivity and uncertainty analysis}. Chapman Hall/CRC,
  London.

\bibitem[{Chadwick et~al.(2011)Chadwick, Herman, Oblo{\v z}insk{\'y}, Dunn,
  Danon, Kahler, Smith, Pritychenko, Arbanas, Arcilla, Brewer, Brown, Capote,
  Carlson, Cho, Derrien, Guber, Hale, Hoblit, Holloway, Johnson, Kawano,
  Kiedrowski, Kim, Kunieda, Larson, Leal, Lestone, Little, {McCutchan},
  {MacFarlane}, {MacInnes}, Mattoon, {McKnight}, Mughabghab, Nobre, Palmiotti,
  Palumbo, Pigni, Pronyaev, Sayer, Sonzogni, Summers, Talou, Thompson, Trkov,
  Vogt, {van der Marck}, Wallner, White, Wiarda, and Young}]{endfb71}
Chadwick, M.~B., Herman, M., Oblo{\v z}insk{\'y}, P., Dunn, M.~E., Danon, Y.,
  Kahler, A.~C., Smith, D.~L., Pritychenko, B., Arbanas, G., Arcilla, R.,
  Brewer, R., Brown, D.~A., Capote, R., Carlson, A.~D., Cho, Y.~S., Derrien,
  H., Guber, K., Hale, G.~M., Hoblit, S., Holloway, S., Johnson, T.~D., Kawano,
  T., Kiedrowski, B.~C., Kim, H., Kunieda, S., Larson, N.~M., Leal, L.,
  Lestone, J.~P., Little, R.~C., {McCutchan}, E.~A., {MacFarlane}, R.~E.,
  {MacInnes}, M., Mattoon, C.~M., {McKnight}, R.~D., Mughabghab, S.~F., Nobre,
  G. P.~A., Palmiotti, G., Palumbo, A., Pigni, M.~T., Pronyaev, V.~G., Sayer,
  R.~O., Sonzogni, A.~A., Summers, N.~C., Talou, P., Thompson, I.~J., Trkov,
  A., Vogt, R.~L., {van der Marck}, S.~C., Wallner, A., White, M.~C., Wiarda,
  D., Young, P.~G., Dec 2011. {ENDF/B-VII.1} nuclear data for science and
  technology: cross sections, covariances, fission product yields and decay
  data. Nuclear Data Sheets 112~(12), 2887--2996.

\bibitem[{Chadwick et~al.(2006)Chadwick, Oblo{\v z}insk{\'y}, Herman, Greene,
  {McKnight}, Smith, Young, {MacFarlane}, Hale, Frankle, Kahler, Kawano,
  Little, Madland, Moller, Mosteller, Page, Talou, Trellue, White, Wilson,
  Arcilla, Dunford, Mughabghab, Pritychenko, Rochman, Sonzogni, Lubitz,
  Trumbull, Weinman, Brown, Cullen, Heinrichs, {McNabb}, Derrien, Dunn, Larson,
  Leal, Carlson, Block, Briggs, Cheng, Huria, Zerkle, Kozier, Courcelle,
  Pronyaev, and {van der Marck}}]{endfb70}
Chadwick, M.~B., Oblo{\v z}insk{\'y}, P., Herman, M., Greene, N.~M.,
  {McKnight}, R.~D., Smith, D.~L., Young, P.~G., {MacFarlane}, R.~E., Hale,
  G.~M., Frankle, S.~C., Kahler, A.~C., Kawano, T., Little, R.~C., Madland,
  D.~G., Moller, P., Mosteller, R.~D., Page, P.~R., Talou, P., Trellue, H.,
  White, M.~C., Wilson, W.~B., Arcilla, R., Dunford, C.~L., Mughabghab, S.~F.,
  Pritychenko, B., Rochman, D., Sonzogni, A.~A., Lubitz, C.~R., Trumbull,
  T.~H., Weinman, J.~P., Brown, D.~A., Cullen, D.~E., Heinrichs, D.~P.,
  {McNabb}, D.~P., Derrien, H., Dunn, M.~E., Larson, N.~M., Leal, L.~C.,
  Carlson, A.~D., Block, R.~C., Briggs, J.~B., Cheng, E.~T., Huria, H.~C.,
  Zerkle, M.~L., Kozier, K.~S., Courcelle, A., Pronyaev, V., {van der Marck},
  S.~C., Dec 2006. {ENDF/B-VII.0:} next generation evaluated nuclear data
  library for nuclear science and technology. Nuclear Data Sheets 107~(12),
  2931--3060.

\bibitem[{{CSEWG}(2013)}]{endf6_format}
{CSEWG}, Jul. 2013. {ENDF-6 Formats Manual. Data Formats and Procedures for the
  Evaluated Nuclear Data Files ENDF/B-VI and ENDF/B-VII}. {CSEWG} Document
  {ENDF-102} {BNL-90365-2009} Rev.1, Brookhaven National Laboratory, Upton,
  USA.

\bibitem[{{CSEWG-Collaboration}(2001)}]{endfb68}
{CSEWG-Collaboration}, 2001. {Evaluated Nuclear Data File ENDF/B-VI.8}.
  {www.nndc.bnl.gov/endf released in October 2001}.

\bibitem[{D\'iez et~al.(2014)D\'iez, Cabellos, and
  Mart\'inez}]{cj_andes_nd2013}
D\'iez, C.~J., Cabellos, O., Mart\'inez, J.~S., 2014. {Impact of Nuclear Data
  Uncertainties on Advanced Fuel Cycles and their Irradiated Fuel -- a
  Comparison between Libraries}. Nuclear Data Sheets 118~(0), 538 -- 541.
\newline\urlprefix\url{http://www.sciencedirect.com/science/article/pii/S0090375214001586}

\bibitem[{Fiorito et~al.(2014)Fiorito, D\'iez, Cabellos, Stankovskiy, {Van den
  Eynde}, and Labeau}]{luca_fy_paper}
Fiorito, L., D\'iez, C.~J., Cabellos, O., Stankovskiy, A., {Van den Eynde}, G.,
  Labeau, P.~E., 2014. {Fission yield covariance generation and uncertainty
  propagation through fission pulse decay heat calculation}. Annals of Nuclear
  Energy 69~(0), 331--343.
\newline\urlprefix\url{http://www.sciencedirect.com/science/article/pii/S0306454914000565}

\bibitem[{Garc\'ia-Herranz et~al.(2008)Garc\'ia-Herranz, Cabellos, Sanz, Juan,
  and Kuijper}]{hybrid_method}
Garc\'ia-Herranz, N., Cabellos, O., Sanz, J., Juan, J., Kuijper, J.~C., 2008.
  {Propagation of statistical and nuclear data uncertainties in Monte Carlo
  burn-up calculations}. Annals of Nuclear Energy 35~(4), 714--730.
\newline\urlprefix\url{http://www.sciencedirect.com/science/article/pii/S0306454907001958}

\bibitem[{Herman et~al.(2011)Herman, Oblo{\v z}insk{\'y}, Mattoon, Pigni,
  Hoblit, Mughabghab, Sonzogni, Talou, Chadwick, Hale, Kahler, Kawano, Little,
  and Young}]{commara20}
Herman, M.~W., Oblo{\v z}insk{\'y}, P., Mattoon, C.~M., Pigni, M., Hoblit, S.,
  Mughabghab, S.~F., Sonzogni, A.~A., Talou, P., Chadwick, M.~B., Hale, G.~M.,
  Kahler, A.~C., Kawano, T., Little, R.~C., Young, P.~G., 2011. {COMMARA-2.0
  neutron cross section covariance library}. Tech. Rep. BNL-94830-2011, BNL.

\bibitem[{Holmes and Hawari(2014)}]{Sab_usa}
Holmes, J.~C., Hawari, A.~I., 2014. {Generation of an Covariance Matrix by
  Monte Carlo Sampling of the Phonon Frequency Spectrum}. Nuclear Data Sheets
  118~(0), 392 -- 395.
\newline\urlprefix\url{http://www.sciencedirect.com/science/article/pii/S0090375214001197}

\bibitem[{Ivanov et~al.(2012)Ivanov, Avramova, Kamerow, Kodeli, Sartori,
  Ivanov, and Cabellos}]{uam_specification}
Ivanov, K., Avramova, M., Kamerow, S., Kodeli, I., Sartori, E., Ivanov, E.,
  Cabellos, O., 2012. {Benchmark for Uncertainty Analysis in Modeling (UAM) for
  Design, Operation and Safety Analysis of LWRs}. NEA/NSC/DOC(2012) I.

\bibitem[{{Joint Evaluated Fission and Fusion File (JEFF)
  project}({2014})}]{jeff32}
{Joint Evaluated Fission and Fusion File (JEFF) project}, March {2014}.
  {JEFF-3.2 evaluated data library - Neutron data}.
\newline\urlprefix\url{{http://www.oecd-nea.org/dbforms/data/eva/evatapes/jeff\_32/}}

\bibitem[{Koning and Rochman(2008)}]{tmc1}
Koning, A.~J., Rochman, D., 2008. {Towards sustainable nuclear energy: Putting
  nuclear physics to work}. Ann. Nucl. Energy 35~(11), 2024--2030.

\bibitem[{Lepp\"anen(2007)}]{serpent}
Lepp\"anen, J., 2007. {Development of A New Monte Carlo Reactor Physics Code}.
  Ph.D. thesis, {Helsinki University of Technology}.

\bibitem[{Little et~al.(2008)Little, Kawano, Hale, Pigni, Herman, Oblo{\v
  z}insk{\'y}, Williams, Dunn, Arbanas, Wiarda, McKnight, McKamy, and
  Felty}]{blo_unc_lib}
Little, R.~C., Kawano, T., Hale, G.~D., Pigni, M.~T., Herman, M., Oblo{\v
  z}insk{\'y}, P., Williams, M.~L., Dunn, M.~E., Arbanas, G., Wiarda, D.,
  McKnight, R.~D., McKamy, J.~N., Felty, J.~R., 2008. {Low-fidelity Covariance
  Project}. Nuclear Data Sheets 109~(12), 2828--2833.
\newline\urlprefix\url{http://www.sciencedirect.com/science/article/pii/S0090375208001063}

\bibitem[{MacFarlane and Kahler(2010)}]{njoy}
MacFarlane, R.~E., Kahler, A.~C., 2010. {Methods for processing ENDF/B-VII with
  NJOY}. Nuclear Data Sheets 111~(12), 2739 -- 2890.
\newline\urlprefix\url{http://www.sciencedirect.com/science/article/pii/S0090375210001006}

\bibitem[{Mart\'inez et~al.(2014)Mart\'inez, Zwermann, Gallner, Puente-Espel,
  Cabellos, Velkov, and Hannstein}]{jsg_nd2013}
Mart\'inez, J.~S., Zwermann, W., Gallner, L., Puente-Espel, F., Cabellos, O.,
  Velkov, K., Hannstein, V., 2014. {Propagation of Neutron Cross Section,
  Fission Yield, and Decay Data Uncertainties in Depletion Calculations}.
  Nuclear Data Sheets 118~(0), 480 -- 483.
\newline\urlprefix\url{http://www.sciencedirect.com/science/article/pii/S0090375214001422}

\bibitem[{Mattoon et~al.(2012)Mattoon, Brown, and Elliott}]{kiwi_tool}
Mattoon, C.~M., Brown, D., Elliott, J.~B., 2012. {Covariance Applications with
  Kiwi}. EPJ Web of Conferences 27, 00002.
\newline\urlprefix\url{http://dx.doi.org/10.1051/epjconf/20122700002}

\bibitem[{{Oak Ridge National Laboratory}(2009)}]{scale60}
{Oak Ridge National Laboratory}, January 2009. {SCALE: A Modular Code System
  for Performing Standardized Computer Analysis for Licensing Evaluation,
  ORNL/TM-2005/39, Version 6.0}. Oak Ridge, Tennessee, USA.

\bibitem[{Pelowitz(2005)}]{mcnpx}
Pelowitz, D.~B., Apr. 2005. {MCNPX User's Manual, Version 2.5.0,
  LA-CP-05-0369}. Los Alamos, New Mexico, USA.

\bibitem[{Poston and Trellue(1998)}]{monteburns}
Poston, D.~I., Trellue, H.~R., 1998. {User`s manual, version 1.00 for
  Monteburns, version 3.01}. Tech. Rep. LA-UR--98-2718, LANL.

\bibitem[{Rochman and Koning(2012)}]{Sab_nrg}
Rochman, D., Koning, A.~J., 2012. {Random Adjustment of the H in H$_2$O Neutron
  Thermal Scattering Data}. Nuclear Science and Engineering 172~(3), 287--299.

\bibitem[{Rochman et~al.(2012)Rochman, Koning, and {Da
  Cruz}}]{nrg_tmc_pwr_burnup_pincell}
Rochman, D., Koning, A.~J., {Da Cruz}, D.~F., September 2012. {Propagation of
  $^{235,236,238}$U and $^{239}$Pu Nuclear Data Uncertainties for a Typical PWR
  Fuel Element}. Nuclear Technology 179~(3), 323--338.

\bibitem[{Rochman and Sciolla(2012)}]{nrg_tmc_pwr_burnup_pincell_report}
Rochman, D., Sciolla, C.~M., April 2012. {"Total Monte Carlo" Uncertainty
  propagation applied to the Phase I-1 burnup calculation}. Tech. Rep. NRG
  Report 113696, Nuclear Research and Consultancy Group, Peten, The
  Netherlands.
\newline\urlprefix\url{{ftp://ftp.nrg.eu/pub/www/talys/bib_rochman/tmc.nrg.pdf}}

\bibitem[{Rochman et~al.(2014)Rochman, Zwermann, van~der Marck, Koning,
  Sj\"ostrand, Helgesson, and Krzykacz-Hausmann}]{fastTMC2013}
Rochman, D., Zwermann, W., van~der Marck, S.~C., Koning, A.~J., Sj\"ostrand,
  H., Helgesson, P., Krzykacz-Hausmann, B., Jul. 2014. {Efficient use of Monte
  Carlo: uncertainty propagation}. Nuclear Science and Engineering 177~(3),
  337--349.

\bibitem[{Sabouri et~al.(2014)Sabouri, Bidaud, Dabiran, Lecarpentier, and
  Ferragut}]{sabouri_nd2013}
Sabouri, P., Bidaud, A., Dabiran, S., Lecarpentier, D., Ferragut, F., 2014.
  {Propagation of Nuclear Data Uncertainties in Deterministic Calculations:
  Application of Generalized Perturbation Theory and the Total Monte Carlo
  Method to a PWR Burnup Pin-Cell}. Nuclear Data Sheets 118~(0), 523 -- 526.
\newline\urlprefix\url{http://www.sciencedirect.com/science/article/pii/S0090375214001549}

\bibitem[{Salvatores et~al.(2008)Salvatores, Aliberti, Dunn, Hogenbirk,
  Ignatyuk, Ishikawa, Kodeli, Koning, McKnight, Mills, Oblo{\v z}insk{\'y},
  Palmiotti, Plompen, Rimpault, Rugama, Talou, and Yang}]{wpec_sg_26}
Salvatores, M., Aliberti, G., Dunn, M., Hogenbirk, A., Ignatyuk, A., Ishikawa,
  M., Kodeli, I., Koning, A.~J., McKnight, R., Mills, R.~W., Oblo{\v
  z}insk{\'y}, P., Palmiotti, G., Plompen, A., Rimpault, G., Rugama, Y., Talou,
  P., Yang, W.~S., 2008. {Uncertainty and Target Accuracy Assessment for
  Innovative Systems Using Recent Covariance Data Evaluations}. Tech. Rep.
  Report NEA/WPEC-26, NEA.

\bibitem[{Sanz et~al.(2008)Sanz, Cabellos, and Garc\'ia-Herranz}]{acab_manual}
Sanz, J., Cabellos, O., Garc\'ia-Herranz, N., Dec. 2008. {ACAB Inventory code
  for nuclear applications: User's Manual V. 2008}. Madrid, Spain.

\bibitem[{Shibata et~al.(2011)Shibata, Iwamoto, Nakagawa, Iwamoto, Ichihara,
  Kunieda, Chiba, Furutaka, Otuka, Ohasawa, Murata, Matsunobu, Zukeran, Kamada,
  and Katakura}]{jendl4}
Shibata, K., Iwamoto, O., Nakagawa, T., Iwamoto, N., Ichihara, A., Kunieda, S.,
  Chiba, S., Furutaka, K., Otuka, N., Ohasawa, T., Murata, T., Matsunobu, H.,
  Zukeran, A., Kamada, S., Katakura, J.-I., 2011. {JENDL-4.0: A New Library for
  Nuclear Science and Engineering}. Journal of Nuclear Science and Technology
  48~(1), 1--30.

\bibitem[{Shibata et~al.(2002)Shibata, Kawano, Nakagawa, Iwamoto, Katakura,
  Fukahori, Chiba, Hasegawa, Murata, Matsunobu, Ohsawa, Nakajima, Yoshida,
  Zukeran, Kawai, Baba, Ishikawa, Asami, Watanabe, Watanabe, Igashira,
  Yamamuro, Kitazawa, Yamano, and Takano}]{jendl33}
Shibata, K., Kawano, T., Nakagawa, T., Iwamoto, O., Katakura, J.-I., Fukahori,
  T., Chiba, S., Hasegawa, A., Murata, T., Matsunobu, H., Ohsawa, T., Nakajima,
  Y., Yoshida, T., Zukeran, A., Kawai, M., Baba, M., Ishikawa, M., Asami, T.,
  Watanabe, T., Watanabe, Y., Igashira, M., Yamamuro, N., Kitazawa, H., Yamano,
  N., Takano, H., 2002. {Japanese Evaluated Nuclear Data Library Version 3
  Revision-3: JENDL-3.3}. Journal of Nuclear Science and Technology 39~(11),
  1125--1136.

\bibitem[{Smith(2011)}]{endfb70_enfb71_cov_changes}
Smith, D.~L., 2011. {Evaluated Nuclear Data Covariances: The Journey From
  ENDF/B-VII.0 to ENDF/B-VII.1}. Nuclear Data Sheets 112~(12), 3037--3053.
\newline\urlprefix\url{http://www.sciencedirect.com/science/article/pii/S0090375211001153}

\bibitem[{{Studsvik Scandpower}(2010)}]{CASMO5}
{Studsvik Scandpower}, Jul. 2010. {CASMO-5/CASMO-5M - A Fuel Assembly Burnup
  Program User's Manual, SSP-07/431, Rev. 1}.

\bibitem[{Sublet et~al.(2012)Sublet, Eastwood, and Morgan}]{fispact2}
Sublet, J.-C., Eastwood, J.~W., Morgan, J.~G., 2012. {The FISPACT-II User
  Manual}. Tech. Rep. CCFE-R(11)11 Issue 3, {CCFE}.

\bibitem[{Venard et~al.(2009)Venard, Santamarina, Leclainche, and
  Mounier}]{rib_tool}
Venard, C., Santamarina, A., Leclainche, A., Mounier, C., 2009. {The RIB tool
  for the determination of computational bias and associated uncertainty in the
  CRISTAL criticality-safety package}. In: {2009 Nuclear Criticality Safety
  Division (NCSD 2009) Topical Meeting on Realism, Robustness and the Nuclear
  Renaissance, Richland, WA (United States), 13-17 Sep 2009}.

\bibitem[{Wiarda and Dunn(2008)}]{puff4}
Wiarda, D., Dunn, M.~E., 2008. {PUFF-IV: A Code for Processing ENDF Uncertainty
  Data into Multigroup Covariance Matrices}. Tech. Rep. ORNL/TM-2006/147/R1,
  ORNL.

\bibitem[{Wieselquist et~al.(2013)Wieselquist, Zhu, Vasiliev, and
  Ferroukhi}]{PSI}
Wieselquist, W., Zhu, T., Vasiliev, A., Ferroukhi, H., 2013. {PSI Methodologies
  for Nuclear Data Uncertainty Propagation with CASMO-5M and MCNPX: Results for
  OECD/NEA UAM Benchmark Phase I}. Science and Technology of Nuclear
  Installations 2013.

\bibitem[{{X-5 Monte Carlo Team}(2003)}]{mcnp5}
{X-5 Monte Carlo Team}, Apr. 2003. {MCNP -- A General Monte Carlo N-Particle
  Transport Code, LA-CP-03-0245, Version 5}. Los Alamos, New Mexico, USA.

\bibitem[{Zhu et~al.(2014)Zhu, Rochman, Vasiliev, Ferroukhi, Wieselquist, and
  Pautz}]{PSI_nuss}
Zhu, T., Rochman, D., Vasiliev, A., Ferroukhi, H., Wieselquist, W., Pautz, A.,
  2014. {Comparison of Two Approaches for Nuclear Data Uncertainty Propagation
  in MCNPX for Selected Fast Spectrum Critical Benchmarks}. Nuclear Data Sheets
  118~(0), 388 -- 391.
\newline\urlprefix\url{http://www.sciencedirect.com/science/article/pii/S0090375214001185}

\bibitem[{Zwermann et~al.(2009)Zwermann, Krzykacz-Hausmann, Gallner, and
  Pautz}]{xsusa}
Zwermann, W., Krzykacz-Hausmann, B., Gallner, L., Pautz, A., 29 Sep. - 2 Oct
  2009. {Influence of Nuclear Covariance Data on Reactor Core Calculations}.
  In: Proc. Second International Workshop on Nuclear Data Evaluation for
  Reactor Applications (WONDER 2009), Cadarache (France). pp. 99--104.

\bibitem[{Zwermann et~al.(2012)Zwermann, Krzykacz-Hausmann, Gallner, Pautz, and
  Velkov}]{fastGRS2012}
Zwermann, W., Krzykacz-Hausmann, B., Gallner, L., Pautz, A., Velkov, K., 2012.
  {Aleatoric and epistemic uncertainties in sampling based nuclear data
  uncertainty and sensitivity analyses}. In: {Int. Conf. PHYSOR 2012:
  Conference on Advances in Reactor Physics - Linking Research, Industry, and
  Education, Knoxville, TN (United States), 15-20 Apr 2012}.
\newline\urlprefix\url{{http://www.osti.gov/scitech/servlets/purl/22105777}}

\end{thebibliography}







\end{document}